\documentclass[10pt]{article}
\usepackage{bbm}
\usepackage[final]{graphics}
\usepackage{amsmath}
\usepackage{amsfonts,amsbsy}
\usepackage{amssymb}
\def\empile#1\over#2{\mathrel{\mathop{\kern 0pt#1}\limits_{#2}}}
\def\bs{\boldsymbol}

\newcommand{\spc}{\mbox{ }}
\def\p{{\boldsymbol p}}

\def\l{{\boldsymbol l}}
\def\k{{\boldsymbol k}}

\def\x{{\boldsymbol x}}

\def\v{{\boldsymbol v}}

\def\u{{\boldsymbol u}}

\def\d3p{\frac{d^3\p}{(2\pi)^3}E_\p}

\def\bfk{{\bf k}} 
 
\def\bfx{{\bf x}}

\catcode`\@=11

\newcount\@tempcntc
\def\@citex[#1]#2{\if@filesw\immediate\write\@auxout{\string\citation{#2}}\fi
  \@tempcnta\z@\@tempcntb\m@ne\def\@citea{}\@cite{%
        \@for\@citeb:=#2\do%
    {\@ifundefined{b@\@citeb}%
        {\@citeo\@tempcntb\m@ne\@citea%
                \def\@citea{,\penalty\@m\ }{\bf ?}\@warning%
                {Citation `\@citeb' on page \thepage \space undefined}}%
        {\setbox\z@\hbox{\global\@tempcntc0\csname b@\@citeb\endcsname\relax}
     \ifnum\@tempcntc=\z@ \@citeo\@tempcntb\m@ne%
       \@citea\def\@citea{,\penalty\@m}%
       \hbox{\csname b@\@citeb\endcsname}%
     \else%
      \advance\@tempcntb\@ne%
      \ifnum\@tempcntb=\@tempcntc%
      \else\advance\@tempcntb\m@ne\@citeo%
      \@tempcnta\@tempcntc\@tempcntb\@tempcntc\fi\fi}}\@citeo}{#1}}%

\def\@citeo{\ifnum\@tempcnta>\@tempcntb\else\@citea
  \def\@citea{,\penalty\@m}%
  \ifnum\@tempcnta=\@tempcntb\the\@tempcnta\else
   {\advance\@tempcnta\@ne\ifnum\@tempcnta=\@tempcntb \else
\def\@citea{--}\fi
    \advance\@tempcnta\m@ne\the\@tempcnta\@citea\the\@tempcntb}\fi\fi}

\catcode`\@=12

\begin{document}

\title{\bf The initial spectrum of fluctuations\\ in the little bang}
\author{Kevin Dusling${}^{(1)}$, Fran\c cois Gelis${}^{(2)}$, Raju Venugopalan${}^{(3)}$}

\maketitle

\begin{enumerate}
\item Physics Department, North Carolina State University,\\
   Raleigh, NC 27695, USA
\item Institut de Physique Th\'eorique (URA 2306 du CNRS),
  CEA/DSM/Saclay,\\  91191, Gif-sur-Yvette Cedex, France
\item Physics Department, Bldg. 510A, Brookhaven National Laboratory,\\
   Upton, NY 11973, USA
\end{enumerate}

\begin{abstract}
  High parton densities in ultra-relativistic nuclear collisions
  suggest a description of these collisions wherein the high energy
  nuclear wavefunctions and the initial stages of the nuclear
  collision are dominated by classical fields. This underlying
  paradigm can be significantly improved by including quantum
  fluctuations around the classical background fields. One class of
  these contributes to the energy evolution of multi-parton
  correlators in the nuclear wavefunctions. Another dominant class of
  unstable quantum fluctuations grow rapidly with proper time $\tau$
  after the collision. These {\sl secular} terms appear at each loop
  order; the leading contributions can be resummed to all loop orders
  to obtain expressions for final state observables. The all-order
  result can be expressed in terms of the spectrum of fluctuations on
  the initial proper time surface. We compute, in $A^\tau=0$ gauge,
  the essential elements in this fluctuation spectrum--the small
  quantum fluctuation modes in the classical background field.  With
  our derivation in QCD, we have all the ingredients to compute
  inclusive quantities in heavy ion collisions at early times
  including i) all--order leading logs in Bjorken $x_{1,2}$ of the two
  nuclei, ii) all strong multiple scattering contributions, and iii)
  all--order leading secular terms.  In the simpler analogous
  formalism for a scalar $\phi^4$ theory, numerical analysis of the
  behavior of the energy-momentum tensor is strongly suggestive of
  early hydrodynamic flow in the system~\cite{DusliEGV1}. In QCD, in
  addition to studying the possible early onset of hydrodynamic
  behavior, additional important applications of our results include
  a) the computation of sphaleron transitions off-equilibrium, and b)
  ``jet quenching'', or medium modification of parton spectra, in
  strong color fields at early times.
\end{abstract}


\section{Introduction}

The large flow measured in heavy ion collisions at
RHIC~\cite{Adamsa3,Adcoxa1,Arsena2,Backa2} and more recently at the
LHC~\cite{Aamoda1} can be described in hydrodynamic models that have
both a nearly {\sl perfect fluid} value of the shear viscosity to
entropy ratio of the quark-gluon matter produced and fairly short
thermalization times that usually range between 0.5 and 2
fermis/c~\cite{LuzumR1,Romat1,Teane1} (depending on the assumptions
made about the initial conditions and the implementation of the
freeze-out). How isotropization and (subsequently) thermalization is
achieved in heavy ion collisions is an outstanding problem which
requires that one understand the properties of the relevant degrees of
freedom in the nuclear wavefunctions and how these degrees of freedom
are released in a collision to produce quark-gluon matter. An {\sl ab
  initio} approach to the problem can be formulated within the Color
Glass Condensate (CGC) effective field theory, which describes the
relevant degrees of freedom in the nuclei as dynamical gauge fields
coupled to static color sources~\cite{GelisIJV1,IancuV1}. The
computational power of this effective theory is a consequence of the
dynamical generation of semi-hard {\sl saturation
  scale}~\cite{GriboLR1,MuellQ1} larger than the intrinsic
non-perturbative QCD scale ($Q_s^2\gg \Lambda_{\rm QCD}^2$), which
allows for a weak coupling treatment of the relevant degrees of
freedom~\cite{McLerV1,McLerV2,McLerV3} in the high energy nuclear
wavefunctions.

There has been significant recent progress in applying the CGC
effective field theory to studying the early time behavior of the
quark-gluon matter called {\sl Glasma}~\cite{LappiM1} produced in the
initial {\sl little bang} of a high energy heavy ion
collision. Inclusive quantities such as the pressure and the energy
density in the Glasma can be written as expressions that factorize, to
leading logarithmic accuracy in the longitudinal momentum fraction
$x$, the universal properties of the nuclear wavefunctions (measurable
for instance in proton-nucleus or electron-nucleus collisions) from
the final state evolution of the matter in the
collision~\cite{GelisLV3,GelisLV4,GelisLV5}. Key to this approach are
the quantum fluctuations around the classical fields. In particular,
quantum fluctuations that are invariant under boosts can be shown to
factorize into universal density functionals that encode the
multi-parton correlations in the nuclear wavefunctions. The evolution
of these density functionals with energy is described by the JIMWLK
renormalization group
equation~\cite{JalilKMW1,JalilKLW1,JalilKLW2,JalilKLW3,JalilKLW4,IancuLM1,IancuLM2,FerreILM1}.

There are however quantum fluctuations that are not boost
invariant. It was observed in~\cite{RomatV1,RomatV2,RomatV3}, via
numerical solutions (see also \cite{FujiiI1,FujiiII1} for a
semi-analytic discussion of some instabilities in the solutions
Yang-Mills equations) of the classical Yang-Mills equations, that
rapidity dependent quantum fluctuations in the expanding Glasma are
unstable and grow exponentially as the square root of the proper time
$\tau$ after the collision. In fact, both the existence and the
specific time dependence of these instabilities was anticipated based
on studies of the Weibel instabilities in expanding anisotropic
Yang-Mills plasmas~\cite{Mrowc3,RebhaRS1,ArnolMY4,MrowcT1}.  The
unstable quantum fluctuations (initially of order ${\cal O}(1)$)
become comparable in size to the classical field (of order ${\cal
  O}(g^{-1})$) on a very short time scale $\tau \sim Q_s^{-1}$.
Fortunately, one can isolate and resum these rapidly growing {\sl
  secular} divergences to all orders in perturbation theory. The
resulting expressions are free of secular divergences, and can be
rephrased as an average over a spectrum of Gaussian fluctuations of
the initial data for the classical field encountered at {\sl leading
  order}.  A similar observation was made previously in the context of
inflationary cosmology~\cite{PolarS1,Son1}.

In a previous paper~\cite{DusliEGV1}, we developed this formalism for
a scalar $\phi^4$ theory which, like QCD, has a dimensionless coupling
in 3+1 dimensions and has unstable modes.  We computed the spectrum of
fluctuations and showed that the resummed expression for the pressure
and energy density can similarly be expressed as an ensemble average
over quantum fluctuations.  The rapid growth of the unstable
fluctuations has drastic consequences.  Without resummation, the
relation between the energy density and the pressure is not single
valued. For the resummed expressions, while the relation between the
pressure and energy density is not single valued at early times, it
becomes so after a finite time evolution. This development of an {\sl
  equation of state} therefore allows one to write the conservation
equation for the resummed energy momentum tensor $T^{\mu\nu}$ as the
closed form set of equations corresponding to the hydrodynamical
evolution of a relativistic fluid. This result can be interpreted as
arising from a phase decoherence of the classical field configurations
with different initial conditions given by the ensemble of quantum
fluctuations.  In this theory, the period of the classical
trajectories is proportional to the amplitude of the field.
Anharmonicity occurs in any non--linear system and we expect the same
to hold for QCD.  As the different trajectories become phase shifted
for different amplitudes there are cancellations resulting in a single
valued relation between the pressure and the energy density.

This phenomenon shares several common features with Srednicki's
hypothesis of {\sl eigenstate thermalization} and {\sl Berry's
  conjecture}~\cite{Berry1,Deuts1,Sredn1,Jarzy1,RigolDO1}. Berry
conjectured in \cite{Berry1} that high lying energy eigenstates of
systems whose classical counterpart is chaotic have very complicated
wavefunctions that for many purposes behave like random
Gaussian\footnote{Naturally, the wavefunction of a given eigenstate is
  not a random function. Berry's conjecture means that for the purpose
  of computing the expectation value of sufficiently inclusive
  observables, one can replace the true wavefunction by a random
  Gaussian function.} functions. A system in such an eigenstate would
display features reminiscent of thermal equilibrium, despite being in
a pure quantum state~\cite{Sredn1}. For a system starting initially in
a coherent state rather than an energy eigenstate, thermalization
would merely amount to losing the initial coherence.  Although these
ideas where formulated in much simpler systems, they may have some
relevance to QCD since here also the underlying classical theory is
believed to be chaotic~\cite{BiroGMT1,HeinzHLMM1}.

In this paper, we shall focus on computing the initial spectrum of
fluctuations in the Glasma formed at early times after a heavy ion
collision. The classical background field at $\tau=0^+$ in the Glasma
can be expressed, from the continuity of the Yang-Mills equations
across the light-cone~\cite{KovneMW1,KovchR1}, in terms of classical
solutions of the Yang-Mills equations for each of the two nuclei
before the collision. For later times, analytical solutions are not
known\footnote{For some interesting recent attempts,
  see~\cite{BlaizM1,BlaizLM1}.}; however, the Yang-Mills equations
have been solved numerically with the initial conditions at
$\tau=0^+$~\cite{KrasnV3,KrasnV1,KrasnNV2,Lappi1,Lappi4}; for a nice
review, see \cite{Lappi6}. Fortunately, inclusive quantities such as
components of the energy-momentum tensor are sensitive only to the
initial spectrum of fluctuations about the classical field at
$\tau=0^+$, which can be calculated analytically. Specifically, we
will solve the small fluctuations equations of motion in $A^\tau=0$
gauge, in order to obtain a complete orthonormal basis of these
fluctuations. There was a first attempt to compute the small
fluctuations 2-point correlator in the Glasma~\cite{FukusGM1} which,
as we shall discuss, was incomplete because it did not include fully
the structure of the background field.

The paper is organized as follows. In the next section, we will
outline the power counting of higher order contributions in the Glasma
and emphasize the necessity of resumming secular terms. We isolate the
leading contributions and obtain an expression for the resummed
leading secular divergences. We show that this expression for
inclusive quantities can be rewritten as a path integral over a
spectrum of fluctuations times the leading order (classical)
expression for the inclusive quantity. The only unknown ingredient in
this reformulation are the small fluctuation fields on the initial
proper time hypersurface. Additional sub-sections discuss gauge
invariance issues and the renormalization of ultraviolet
divergences. In section \ref{sec:spectrum1}, we will show how to
compute the small fluctuation fields in the vacuum. We first obtain an
inner product for fluctuations on a space-like Cauchy surface that we
use to define the orthogonality between a pair of fields. We will
further prove that the inner product is independent of the chosen
surface. We then show that the small fluctuation fields can be
expressed as a linear combination of modes whose coefficients are
Gaussian-distributed random complex numbers. (This is equivalent to
diagonalizing the small fluctuation correlator on the initial proper
time surface.) Because even the computation of the free fluctuations
in $A^\tau=0$ gauge is non-trivial, we shall first solve small
fluctuation equations in the vacuum. Then, in the section
\ref{sec:glasma}, we shall solve the small fluctuation equations in
the background classical field of the Glasma to construct the
corresponding physical small fluctuation modes. Section \ref{sec:algo}
outlines a practical algorithm to compute inclusive quantities
(including all leading logarithms in $x$, and all leading secular
contributions) as a function of proper time. As we shall demonstrate,
the complexity of this space-time evolution is manageable, and amounts
to diagonalizing certain matrices on the initial proper time
surface. In the final section, we re-state our key results and discuss
some important applications. These include a) a systematic study of
possible thermalization of the quantum system whose evolution with
proper time we plan to simulate numerically just as for the scalar
case studied previously. b) The nature and role of sphaleron
transitions in the early time dynamics of the system. c) The medium
modification of hard probes to study 'jet quenching' at early times.
An open question for future work is to explore until what times these
analysis are valid and how one can incorporate sub-leading
contributions that become increasingly important at late times. There
are two appendices. The first discusses technical aspects of the
computation of small fluctuation fields in the vacuum. Expressions for
the Wightman functions for free fields in $A^\tau=0$ gauge are
discussed in the second appendix, where some connections to previous
work on these is also discussed~\cite{FukusGM1,Makhl3}.

\section{Resummation of leading instabilities in the Glasma}
\label{sec:resummation}
We will begin by first outlining how the power counting for computing
inclusive quantities in field theories with strong time dependent
sources is modified due to the presence of secular
divergences. Following this power counting, we derive an explicit
expression for the energy-momentum tensor in heavy ion collisions that
resums the leading instabilities to all loop orders in perturbation
theory. We will show that the resummed expression for the energy
momentum tensor can be expressed as a path integral over the product of two terms. The first is a weight
functional that samples the spectrum of quantum fluctuations on the
initial proper time hypersurface, while the second is the leading order expression
for the energy-momentum tensor. In the latter, the classical field is shifted by
the sampled quantum fluctuations. Computing the initial spectrum of
fluctuations is our primary goal in this paper, the derivation of
which will be discussed at length in sections 3 and 4.  Before we go
there, two further sub-sections will discuss the constraints imposed
by gauge invariance on the spectrum of fluctuations and the nature of
ultraviolet divergences respectively.

\subsection{Power counting of unstable modes in the Glasma}
In previous works~\cite{GelisV2,GelisV3}, it was shown that the
problem of computing leading order (LO) and next-to-leading order
(NLO) contributions to {\sl inclusive} quantities --such as components
of the energy momentum tensor-- in field theories with strong time
dependent sources can be formulated as an initial value problem where
a classical field determined on an initial Cauchy surface is evolved
up to the time at which the (local) observable is computed. Because
one anticipates that a semi-hard scale $Q_s^2\gg \Lambda_{_{\rm
    QCD}}^2$ is generated by the non-linear QCD dynamics at high
energy~\cite{GriboLR1,MuellQ1}, a systematic weak coupling expansion
of these inclusive quantities is feasible.  One can formally arrange
the perturbative expansion of an observable such as the energy
momentum tensor as
\begin{equation}
{\cal O}\left[\rho_1,\rho_2\right]
=
\frac{1}{g^2}\Big[c_0+c_1 g^2+c_2 g^4+\cdots\Big]\; ,
\label{eq:expansion}
\end{equation}
where each term corresponds to a different loop order. Each of the
coefficients $c_n$ is itself an infinite series of terms involving
arbitrary orders in $(g\rho_{1,2})^p$. These terms are all of order
unity because the color charge densities are of order $\rho_{1,2}\sim
{\cal O}(g^{-1})$ in a large nucleus at high energy. The color
charge densities correspond to the large $x$ color sources in either
nucleus 1 or nucleus 2 respectively in a heavy ion collision. Their
evolution with the separation scale between sources and fields is
described by the JIMWLK equation, which will be stated shortly.  The LO
contribution comes from the first coefficient $c_0$,
\begin{equation}
{\cal O}_{_{\rm LO}}[\rho_1,\rho_2]\equiv\frac{c_0}{g^2}\; .
\end{equation}
This leading term $c_0/g^2$ has been studied extensively for the
single inclusive gluon
distribution in A+A collisions~\cite{KrasnV3,KrasnV1,KrasnNV2,Lappi1} and recently for the double
inclusive distribution as well~\cite{LappiSV1}.

Following this terminology, we denote
\begin{equation}
{\cal O}_{_{\rm NLO}}[\rho_1,\rho_2]\equiv c_1\;,\qquad
{\cal O}_{_{\rm NNLO}}[\rho_1,\rho_2]\equiv c_2\,g^2\; ,\quad\cdots
\end{equation}
At each order in the loop expansion, there can arise contributions
from the loop integrals which are of the same magnitude as lower
orders. One set of such contributions are the increasingly large
logarithms of the momentum fractions $x_{1,2}$ of partons in the
nuclear wavefunctions as higher energies, or equivalently smaller
values of $x_{1,2}$, are achieved in nuclear collisions. The term
$c_n$ can have up to $n$ powers of such logarithms, with {\sl leading
logarithmic} terms identified as terms that have as many logarithms
as their order in the loop expansion,
\begin{equation}
{\cal O}_{_{\rm LLog}}[\rho_1,\rho_2]\equiv\frac{1}{g^2}
\sum_{n=0}^\infty
d_{n}\,\Big[g^{2}\,\ln\left(\frac{1}{x_{1,2}}\right)\Big]^n\; ,
\label{eq:LLog}
\end{equation}
where $d_n$ is the coefficient of $n$-th term in the leading log
expansion. We were able to
show~\cite{GelisLV3,GelisLV4,GelisLV5} that the leading
logarithmic contributions in $x_{1,2}$, after averaging over the
sources $\rho_{1,2}$ factorize into the expression
\begin{equation}
\langle {\cal O}\rangle_{_{\rm LLog}}
= \int [D\rho_1 D\rho_2]\;
W_{x_1}[\rho_1]\, W_{x_2}[\rho_2]\;
{\cal O}_{_{\rm LO}}\left[\rho_1,\rho_2\right] \; ,
\label{eq:fact-formula}
\end{equation}
where $W_{x_{1,2}} [\rho_{1,2}]$ are the density functionals we
alluded to previously. These obey the JIMWLK
equation~\cite{JalilKMW1,JalilKLW1,JalilKLW2,JalilKLW3,JalilKLW4,IancuLM1,IancuLM2,FerreILM1}
\begin{equation}
\frac{\partial W_{x_{1,2}}[\rho_{1,2}]}{\partial \ln(1/x_{1,2})} = {\cal H}_{1,2}\; W_{x_{1,2}}[\rho_{1,2}] \; .
\label{eq:H-JIMWLK1}
\end{equation}
Here ${\cal H}_{1,2}$ are the JIMWLK Hamiltonians of the two nuclei;
since their explicit form is not essential to the discussion here, we
will refer the interested reader to ref.~\cite{GelisLV3} for explicit
expressions in our notation. Given an initial condition at some
initial $x_0$ value, the JIMWLK equation describes the evolution in
the nuclear wavefunctions of the multi-parton correlators that
contribute to inclusive observables measured in the final state.

The resummation of quantum corrections arising from logarithms in
$x_{1,2}$, as sketched here, takes into account contributions that are
essential in describing the energy evolution of inclusive observables
in heavy ion collisions. These contributions are zero modes in $\nu$,
the Fourier conjugate of the space-time rapidity $\eta$, and are
localized in rapidity around the wave functions of the incoming
nuclei. There are also quantum fluctuations that are non-zero modes of
$\nu$.  Such contributions, that do not bring leading logs of
$1/x_{1,2}$, cannot be factorized into the evolution of the density
functionals $W_{x_{1,2}}$ in eq.~(\ref{eq:fact-formula}). As shown
in~\cite{RomatV1,RomatV2,RomatV3,FukusG1}, these terms can be unstable
and grow exponentially with the square root of the proper time (equal
to $\tau \equiv \sqrt{2 x^+ x^-}$ in light-cone co-ordinates) for a
system undergoing one dimensional longitudinal expansion. Based on 
these considerations, the expansion we sketched in
eqs.~(\ref{eq:expansion}) and (\ref{eq:LLog}) needs to be modified to
keep track also of quantum fluctuations of amplitude $g \exp(\sqrt{
  \mu\tau})$ (where $\mu$ is a growth rate of order $Q_s$) relative to
the leading term.  This is necessary because the rapid growth of these
unstable modes leads to a break down of the perturbative expansion
when
\begin{equation}
\tau\sim \tau_{\rm max}\equiv \mu^{-1}\ln^2\left(\frac{1}{g}\right)
\label{eq:taumax}
\end{equation}
is reached, the proper time at which 1-loop corrections become as
large as the leading order term. The breakdown of the expansion can be
avoided if one resums these divergent contributions, leading to a
resummed result that is well behaved for $\tau\to+\infty$. Taking into
account both the leading logs in $1/x_{1,2}$ and the leading unstable
contributions, the new expansion reads
\begin{equation} {\cal O}_{{\rm LLog}\atop{\rm
      LInst.}}[\rho_1,\rho_2]\equiv\frac{1}{g^2} \sum_{n=0}^\infty
  g^{2n} \sum_{p+q=n}\tilde{d}_{p,q}
  \ln^p\left(\frac{1}{x_{1,2}}\right)\;e^{2q\sqrt{\mu\tau}}\;
  .
\label{eq:LLog+LInst.}
\end{equation}
Thus far, we have only resummed the $q=0$ sector of this formula,
where the result of the resummation is expressed by the factorized
formula (\ref{eq:fact-formula}).  The two sources of leading quantum
fluctuations at this accuracy can be resummed independently because a
given quantum fluctuation mode cannot be at the same time a zero mode
(that generates logarithms in $x_{1,2}$) and a non-zero mode (that
generates a secular divergence in proper time $\tau$). Naturally, in
higher loop corrections, one loop can bring a log of $1/x_{1,2}$ while
another loop brings a secular divergence. This is why
eq.~(\ref{eq:LLog+LInst.}) has terms with both $p$ and $q$ non-zero
simultaneously. But the independence of the two types of
divergences, based ultimately on considerations of causality, leads us
to expect that the double series of eq.~(\ref{eq:LLog+LInst.}) can be
factorized into a series in $p$ times a series in $q$.

\subsection{All orders resummation of the leading secular terms}
We shall now discuss how resumming the leading secular terms modifies
the expression of eq.~(\ref{eq:fact-formula}).  Albeit our
considerations apply to any inclusive quantity, for specificity, we
shall consider here the energy-momentum tensor.

\subsubsection{Reminder of LO and NLO results}
Let us recall first that at leading order in $g^2$, the
energy-momentum tensor $T^{\mu\nu}_{_{\rm LO}}$ is given by
\begin{equation}
T^{\mu\nu}_{_{\rm LO}}(x)=
\frac{1}{4}g^{\mu\nu}{\cal F}^{\alpha\beta}_a {\cal F}_{a,\alpha\beta}
-{\cal F}^{\mu\alpha}_a {\cal F}^{\nu}_a{}_{\alpha} \; ,
\end{equation}
with the field strength tensor defined as
\begin{equation}
{\cal F}^{\mu\nu}_a=\partial^\mu {\cal A}^\nu_a
-
\partial^\nu {\cal A}^\mu_a
+
gf^{abc}{\cal A}^\mu_b {\cal A}^\nu_c\; ,
\end{equation}
where ${\cal A}^\mu_a$ is the solution of the classical
Yang-Mills\footnote{We have written the Yang-Mills equations in a form
  that involves the adjoint representation of the covariant
  derivative, ${\cal D}_\mu^{ab}=\partial_\mu\delta^{ab}-ig{\cal
    A}_\mu^{ab}$, where ${\cal A}_\mu^{ab}$ is the classical gauge
  potential in the adjoint representation. It is important to
  distinguish the ${\cal A}_\mu^{ab}$'s from the ${\cal A}_\mu^a$'s
  that are the components of the SU(3) element ${\cal A}_\mu$ in its
  decomposition over the generators of the algebra, ${\cal
    A}_\mu\equiv {\cal A}_\mu^a t^a$. The two sets of coefficients are
  related by ${\cal A}_\mu^{bc}=-if^{abc}{\cal A}_\mu^a$, since the
  components of the generators in the adjoint representation are
  $(t^a_{\rm adj})_{bc}=-if^{abc}$.}  equations with sources $\rho_{1,2}$ that
vanishes at $x^0\to -\infty$,
\begin{equation}
{\cal D}^{ab}_\mu{\cal F}^{\mu\nu}_b=\delta^{\nu+}\rho^a_1+\delta^{\nu-}\rho^a_2
\quad,\qquad\lim_{x^0\to-\infty} {\cal A}^\mu_a(x)=0\; .
\label{eq:YM}
\end{equation}

One can then express the NLO contribution to the energy-momentum
tensor, for a given distribution of color sources\footnote{Unless
  specified otherwise, the dependence on $\rho_1,\rho_2$, the color
  charge densities in each of the nuclei, will be implicit in our
  discussion.} and at an arbitrary space-time point, as the action of
a functional operator acting on the LO result~\cite{GelisLV3,GelisV2},
\begin{equation}
T^{\mu\nu}_{_{\rm NLO}}(x)
=
\Big[
\int\limits_{\Sigma} d^3\u\;
\beta\cdot{\mathbbm T}_\u
+\frac{1}{2}\int\limits_{\Sigma} d^3\u\; d^3\v\; 
\sum_{\lambda,a}\int\frac{d^3\k}{(2\pi)^3 2k}
[a_{+\k\lambda a}\cdot{\mathbbm T}_\u][a_{-\k\lambda a}\cdot{\mathbbm T}_\v]
\Big]\,
T^{\mu\nu}_{_{\rm LO}}(x)
\; ,
\label{eq:NLO-1}
\end{equation}
where $\Sigma$ is a Cauchy surface where the initial values of the
classical color field and its derivatives are specified. In this
formula, $\beta$ and $a_{\pm\lambda\k a}$ are small corrections to the
gauge field ${\cal A}^\mu$. More specifically, $\beta(\u)$ is the one
loop correction to the classical field on the surface $\Sigma$ (see
\cite{GelisLV3} for more details).  In applications to heavy ion
collisions, a natural choice of $\Sigma$ is the surface at proper time
$\tau=0^+$, which corresponds physically to times just after the two
nuclei have collided. Though $\tau=0^+$ is the initial surface of
choice, we will at the outset consider a generic space-time
hypersurface. The only constraint on $\Sigma$ is that, for the
forthcoming resummation to be effective, it must be located at times
before the unstable modes have become too large\footnote{The NLO
  expression in eq.~(\ref{eq:NLO-1}) does not depend on the choice of
  $\Sigma$. However, our resummed result will depend on this choice
  since it includes only a subset of the higher loop
  corrections. Provided the surface $\Sigma$ is located in a region
  where the unstable fluctuations are still small, the difference
  between various choices of $\Sigma$ is a small correction. We will
  discuss this point further later in the paper.}.  In the right hand
side of eq.~(\ref{eq:NLO-1}), ${\mathbbm T}_\u$ is the generator of
shifts of the initial data for the classical field on
$\Sigma$. Generically, it reads\footnote{It has dimension of
  (mass)$^2$ because dim[$\frac{\delta}{\delta {\cal A}}$] =
  (mass)$^2$ and dim[$a_{\pm\k\lambda a}$] = (mass)$^0$.}
\begin{equation}
a\cdot {\mathbbm T}_\u 
= 
{a^\mu(\u)} \frac{\delta}{\delta {\cal A}^\mu(\u)}
+
(\partial^\nu a^\mu(\u)) 
\frac{\delta}{\delta (\partial^\nu {\cal A}^\mu(\u))}\; ,
\label{eq:shifts}
\end{equation}
where ${\cal A}^\mu(\u)$ (in curly font and without a time argument)
is the value of the classical field on the initial time surface. Note
that in general, specific gauge conditions and specific choices of the
surface $\Sigma$ reduce the number of terms this operator contains. In
its minimal form, it contains one term for each independent field
component or field derivative component that one must specify in the
initial value problem on $\Sigma$.

Let us now explicit a bit more the fields $a_{\pm \k\lambda a}^\mu$
that appear in eq.~(\ref{eq:NLO-1}). They are small fluctuation fields
about the classical field ${\cal A}^\mu$, that obey the equation of
motion\footnote{To avoid cumbersome notations, we have not written
  explicitly the color indices of the various objects. Here, and
  henceforth, the ${\cal D}$'s and ${\cal F}$ should be understood as
  objects in the adjoint representation. For instance ${\cal
    F}^\nu{}_\mu a^\mu$ means ${\cal F}^\nu_{ab}{}_\mu
  a^\mu_b$. Likewise, ${\cal D}_\mu{\cal D}^\mu a^\nu$ is ${\cal
    D}_{ab,\mu}{\cal D}_{bc}^\mu a_c^\nu$.}
\begin{equation}
{\cal D}_\mu\left({\cal D}^\mu a^\nu-{\cal D}^\nu a^\mu\right)
-ig{\cal F}^\nu{}_\mu a^\mu=0\; ,
\label{eq:eom-fluct}
\end{equation}
and that are specified in the remote past by the boundary condition
\begin{equation}
\lim_ {x^0\to -\infty} a_{\pm \k \lambda a}^\mu(x) =
\varepsilon_{\k\lambda}^\mu \,T^a \,e^{\pm i k\cdot x}\; .
\end{equation}
The $T^a$'s are the SU(3) generators and $\varepsilon_{\k\lambda}^\mu$
is the polarization vector. Thus the labels $\k,\lambda, a$ are
respectively the initial momentum, initial polarization and initial
color of the gauge fluctuation represented by $a_{\pm\k\lambda a}$,
and the sign $\pm$ specifies whether it is a positive or negative
energy wave in the remote past. 

\subsubsection{Power counting rules}
At leading order (tree level), the energy momentum tensor is of order
$Q_s^4/g^2$. In the absence of secular divergences, from the power counting described previously, the NLO
corrections should be of order $Q_s^4$. This power counting could be obtained
in eq.~(\ref{eq:NLO-1}) by noting that
\begin{eqnarray}
a_{\pm\k\lambda a} &\sim& {\cal O}(1)\; ,\\
\nonumber\\
\beta&\sim& {\cal O}(g)\; ,
\\
\nonumber\\
{\mathbbm T}_\u &\sim& \frac{\delta}{\delta {\cal A}} \sim {\cal O}(g)
\; ,
\end{eqnarray}
since ${\cal A} \sim {\cal O}(g^{-1})$. The existence of instabilities
implies that we must alter our estimate of the order of magnitude of
the operators ${\mathbbm T}_\u$.  Indeed, since ${\mathbbm T}_\u {\cal
  A}(\tau,\x)$ is the propagator of a small fluctuation over the
background field between a point on the initial proper time surface
and the point $(\tau,\x)$, it grows at the same pace as the
unstable fluctuations. Thus the counting rule for ${\mathbbm T}_\u$
should be modified to read
\begin{equation}
{\mathbbm T}_\u \sim {\cal O}(g\, e^{\sqrt{\mu\tau}})\; .
\end{equation}
The combination ${\mathbbm T}_\u{\mathbbm T}_\v$ in
eq.~(\ref{eq:NLO-1}) then grows as $g^2 \exp(2\sqrt{ \mu\tau})$ which
leads to a break down of the power counting at the proper time
$\tau_{\rm max}$ defined in eq.~(\ref{eq:taumax}). At $\tau_{\rm max}$,
the 1-loop correction becomes as large as the leading order
contribution, and one may anticipate that an infinite series of higher
loop corrections also become equally important at this time.

\subsubsection{Selection of the leading terms}
\begin{figure}[htbp]
\begin{center}
\hfil
\resizebox*{!}{5cm}{\includegraphics{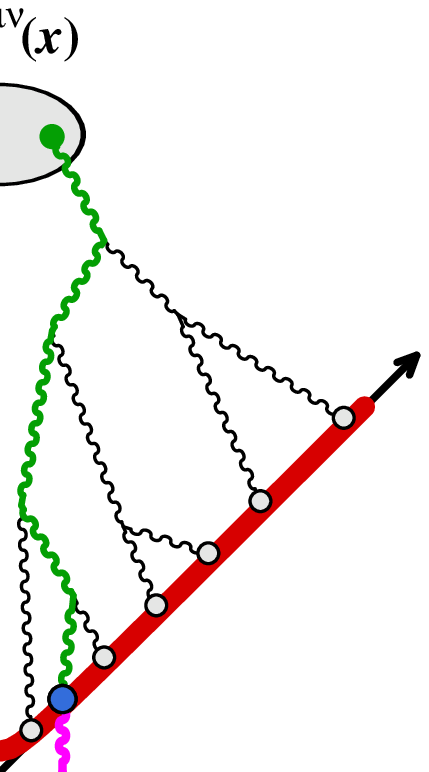}}
\hfil
\resizebox*{!}{5cm}{\includegraphics{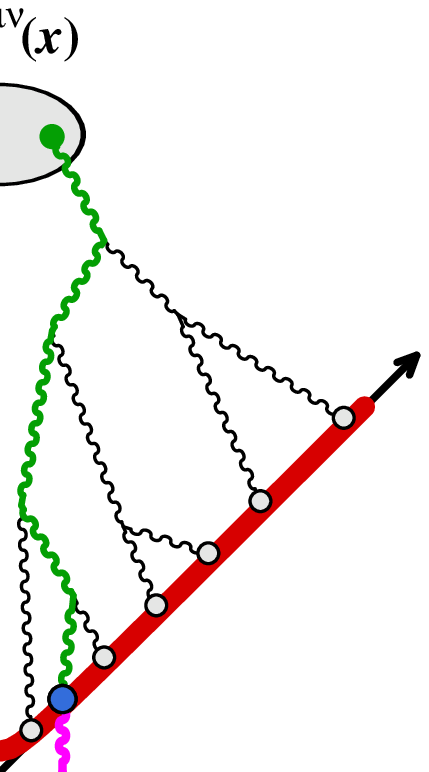}}
\hfil
\end{center}
\caption{\label{fig:2loop}Representation of two examples of 2-loop
  contributions. The thick red line is the $\tau=0^+$ surface on which
  the initial value problem is set up. The open circles represent the
  initial data. The filled blue circles represent operators ${\mathbbm
    T}_{\u,\v}$. Left: contribution with a $\Gamma_4$ that factorizes
  into two $\Gamma_2$'s. Right: contribution with a $\Gamma_3$.}
\end{figure}

Our goal is now to collect from  higher orders all the terms that
are leading at the time $\tau_{\rm max}$. This comprises all the terms
where the extra powers of $g^2$ are compensated by an equal number of
factors of $e^{2\sqrt{\mu\tau}}$.  We presume that a typical higher order
correction to the energy momentum tensor can still be written in the
form of eq.~(\ref{eq:NLO-1}), but with a more general operator acting
on $T^{\mu\nu}_{_{\rm LO}}(x)$ of the form
\begin{equation}
\int\limits_{\Sigma} d^3 {\u_1}\cdots d^3 {\u_n}\;
{\Gamma}_n(\u_1,\cdots,\u_n)\cdot{\mathbbm T}_{\u_1}\cdots {\mathbbm T}_{\u_n}\; .
\label{eq:higher-order}
\end{equation}
\begin{figure}[htbp]
\begin{center}
\resizebox*{!}{5cm}{\includegraphics{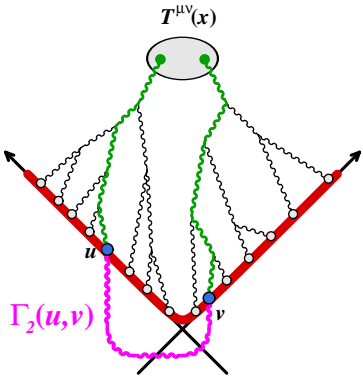}}
\end{center}
\caption{\label{fig:1loop}Representation of the 1-loop contribution
  involving the function $\Gamma_2(\u,\v)$. The thick red line is the
  $\tau=0^+$ surface on which the initial value problem is set up. The
  open circles represent the initial data. The filled blue circles
  represent the two operators ${\mathbbm T}_{\u,\v}$, and the U-shaped
  wavy line represented below the light-cone is the function
  $\Gamma_2(\u,\v)$.}
\end{figure}
Here ${\Gamma}_n$ is an $n$-point function, which may or may not be
simply connected.  This expression has not been proven in general but
results from a conjecture that inclusive quantities at all loop orders
can be expressed purely in terms of retarded propagators, thereby
generalizing known results at LO and NLO. While there are specific
examples of loop contributions that have been checked to satisfy this
conjecture, there are in particular nested loops contributions for
which the conjecture is difficult to confirm. In the figures
\ref{fig:1loop} and \ref{fig:2loop}, we illustrate this formula by
some examples of 1-loop and 2-loop contributions.

With the stated assumption implicit in eq.~(\ref{eq:higher-order}),
the following power counting can be established. If
eq.~(\ref{eq:higher-order}) is a piece of a $L$-loop correction to the
energy-momentum tensor, the order $g^p$ and the number $n$ of points
of the function $\Gamma_n$ are related by
\begin{equation}
n+p=2L\; .
\end{equation}
This formula can be checked for the examples of graphs given in the
figures \ref{fig:1loop} and \ref{fig:2loop}. Note that $p=0$ is the
smallest possible value for $p$. Taking into account the effect of the
instabilities (i.e. one power of $\exp(\sqrt{\mu\tau})$ for each of
the $n$ operators ${\mathbbm T}_{\u_i}$), the order of magnitude of a
contribution obtained from eq.~(\ref{eq:higher-order}) is
\begin{equation}
g^p \left[g\, e^{\sqrt{\mu\tau}}\right]^n\; ,
\end{equation}
relative to the leading order contribution.  If we just count naively
the powers of $g$, the power counting would indicate that this
contribution gives a correction of order $g^{2L}$, a decrease by a
factor $g^2$ for each extra loop. However, because of the instability,
by the time $\tau_{\rm max}$ given in eq.~(\ref{eq:taumax}), the
contribution is instead of order $g^p$ and does not depend anymore on
the number $n$ of ${\mathbbm T}$ operators. At the time $\tau_{\rm
  max}$ all the terms with $p=0$, regardless of the number of loops,
are of the same order while all the remaining terms for which $p>0$
are suppressed by additional powers of $g$.  It is therefore natural
to resum all the $p=0$ terms, and to neglect all those with $p>0$ as
giving sub-leading contributions. This implies that the numbers $n$ of
${\mathbbm T}$ operators must be even and equal to $2L$.  Moreover,
since we keep only terms of order $p=0$ in ${\Gamma}_{2L}$, the only
possibility that remains\footnote{Note that tadpole contributions such
  as the term $\beta\cdot{\mathbbm T}_\u$ in eq.~(\ref{eq:NLO-1}) are
  also excluded since $\beta\sim{\cal O}(g)$.} is to construct
${\Gamma}_{2L}$ as a product of $L$ factors ${\Gamma}_2$ (an example
of which is the left diagram of figure \ref{fig:2loop}) because any
non-factorized contribution to ${\Gamma}_{2L}$ requires more powers of
$g$.

\subsubsection{Resummation: formal expression}
Therefore, the leading operator at $L$ loops in
eq.~(\ref{eq:higher-order}) is the $L$-th power of the $2$-point
operator that appears at 1-loop,
\begin{equation}
\frac{1}{L!}\Bigg[
\frac{1}{2}
\int\limits_{\Sigma} d^3 \u \;d^3 \v\;
{\Gamma}_2(\u,\v)\cdot{\mathbbm T}_{\u}{\mathbbm T}_{\v}
\Bigg]^L\; ,
\end{equation}
where
\begin{equation}
\Gamma_2 (\u,\v)\cdot{\mathbbm T}_{\u}{\mathbbm T}_{\v} 
= \sum_{\lambda,a}\int\frac{d^3\k}{(2\pi)^3 2k}
\;
[a_{+\k\lambda a}\cdot{\mathbbm T}_\u][a_{-\k\lambda a}\cdot{\mathbbm T}_\v]
\; .
\label{eq:small-fluctuation-prop}
\end{equation}
The inverse factorial prefactor is a symmetry factor that prevents
multiple counting when the various factors $\Gamma_2$ are
permuted. Summing all the contributions from $L=0$ (leading order) to
$L=+\infty$, we obtain\footnote{Until the conjecture in
  eq.~(\ref{eq:higher-order}) is proved, one may choose to interpret
  eq.~(\ref{eq:all-orders-formal}) as a well motivated ans\"{a}tz
  resulting from the exponentiation of the NLO result.}
\begin{equation}
T^{\mu\nu}_{\rm resummed}(x)
=
\exp \Bigg[
\frac{1}{2}
\int\limits_{\Sigma}
d^3\u\; d^3 \v\;{\Gamma}_2(\u,\v)\cdot{\mathbbm T}_{\u}{\mathbbm T}_{\v}
\Bigg]\; T^{\mu\nu}_{_{\rm LO}}(x)\; .
\label{eq:all-orders-formal}
\end{equation}
Eq.~(\ref{eq:all-orders-formal}) provides an expression that resums
all the leading contributions\footnote{Subleading contributions such
  as the $p=1$ contribution $g\cdot g\exp(\sqrt{\mu\tau})$ are no
  longer sub-leading by times $\tau \sim \mu^{-1}\ln^2(g^{-2})$. This
  time is only slightly larger than the time $\tau_{\rm max}$ at which
  the $p=0$ terms become important. Therefore, including one by one
  the $p=1$ terms on top of an expression that resums the $p=0$ terms
  is bound to fail -- these contributions must be included all at once via
  a resummation. Having this in mind, a more important question is: do
  these contributions ever become important after they have been
  appropriately resummed? A resummation of the $p=1$ secular terms is
  outside the scope of this work, but it is plausible that they can be
  included in our framework by a modification of the distribution
  $F_0[\alpha]$ of the fluctuations at the initial time.  If this is
  the case, then the $p>1$ terms would simply lead to small
  non-gaussianities in the spectrum of fluctuations. However, the is
  at present nothing more than a conjecture, and the results of this
  paper have to be interpreted with this in mind.} at the time
$\tau=\tau_{\rm max}$.  However, this expression is very formal as it
is expressed in terms of functional derivatives with respect to the
initial conditions for the classical color fields and their time
derivatives on the initial proper time surface $\Sigma$. Fortunately,
as we shall see, we can rewrite this result in a form that is much
more transparent both conceptually and for computational purposes.

\subsubsection{Resummation: path integral representation}
We first recall that the operator ${\mathbbm T}_\u$ defined in
eq.~(\ref{eq:shifts}) is the generator of shifts of the initial
conditions at all points $\u$ on the initial proper time surface $\tau
= 0^+$ for the classical fields ${\cal A}^\mu$ and their time
derivatives $\partial_\tau {\cal A}^\mu$, the latter either being
equal to or simply proportional to the corresponding electric fields,
their canonical conjugate momenta. We can therefore write
\begin{equation}
\exp\Bigg[\int_{\Sigma} d^3\u\;[\alpha\cdot{\mathbbm T}_\u]\Bigg]
\;{\cal F}\big[{\mathbbm A}\big]
=
{\cal F}\big[{\mathbbm A}+\alpha\big]
\; ,
\label{eq:shift-operator}
\end{equation}
where ${\mathbbm A} \equiv ({\cal A}, {\cal E})$ denotes collectively
all the components of the initial classical field and their
canonically conjugate momenta on the initial time surface. One should
similarly understand $\alpha$ to denote small perturbations of both
the initial classical field and their canonically conjugate electric
fields.  We next obtain\footnote{An elementary form of this identity,
\begin{equation*}
  e^{\frac{\alpha}{2}\partial_x^2}\,f(x)
  =
  \int_{-\infty}^{+\infty}dz\;
  \frac{e^{-z^2/2\alpha}}{\sqrt{2\pi\alpha}}\,f(x+z)
  \; ,
\end{equation*}
can be proved by performing a Taylor expansion of the exponential on
the left hand side and of $f(x+z)$ on the right hand side of this
expression. From this simple example, one sees that a Gaussian
operator in derivatives is a {\sl smearing operator} that convolutes
the target function with a Gaussian distribution.}
\begin{equation}
\exp \Bigg[
\frac{1}{2}\!\!
\int\limits_{\Sigma}\!\!
d^3{\u}\,d^3 {\v}\;
{\Gamma}_2(\u,\v)\cdot{\mathbbm T}_{\u}{\mathbbm T}_{\v}
\Bigg]\!
=
\!\int\!\! \big[{\cal D}\alpha\big]\,F_0\big[\alpha\big]
\,
\exp\Bigg[\int\limits_{\Sigma} d^3 {\u}\; [\alpha\cdot{\mathbbm T}_\u]\Bigg]
\; ,
\label{eq:Laplace}
\end{equation}
with\footnote{The unwritten constant prefactor, proportional to $[{\rm
    det}(\Gamma_2)]^{-1/2}$, is such that the distribution
  $F_0\big[\alpha\big]$ has an integral over $\alpha$ normalized to unity.}
\begin{equation}
F_0\big[\alpha\big]
\propto
\exp \Bigg[-\frac{1}{2}\!\!
\int\limits_{\Sigma}\!\!
d^3{\u}\,d^3 {\v}\;
\alpha(\u)\;\Gamma_2^{-1}(\u,\v)\;\alpha(\v)
\Bigg]\; .
\label{eq:Finit-def}
\end{equation}
In eq.~(\ref{eq:Laplace}), the functional integration $[D\alpha]$ is
also a shorthand for integrations over all the components of the
perturbation and of its time derivative on the initial surface.

With the identities~(\ref{eq:shift-operator}) and (\ref{eq:Laplace}) in
hand, we can rewrite our formal result in
eq.~(\ref{eq:all-orders-formal}) as
\begin{equation}
T^{\mu\nu}_{\rm resummed}(x) = \int \!\! \big[{\cal D}\alpha\big]\,
F_0\big[\alpha\big]\;\; T_{_{\rm LO}}^{\mu\nu} [{\mathbbm A} + \alpha] (x)\;,
\label{eq:master-formula}
\end{equation}
where the weight functional $F_0$, corresponding to the initial
spectrum of fluctuations, is defined in eq.~(\ref{eq:Finit-def}). This
result is a central expression of our paper and was sketched
previously in \cite{GelisLV3,GelisLV2}. It was also obtained in
previous works for a scalar field theory~\cite{Son1} and a gauge field
theory~\cite{FukusGM1} respectively using different methods. The
expression is quite remarkable because it demonstrates that the
resummation of loop (quantum) corrections that correspond to the most
unstable configurations in a single heavy ion collision event can be
expressed as an average over a Gaussian distributed ensemble of
classical configurations in the Glasma. Note also that, although this
formula was derived here for the energy-momentum tensor, the power
counting that led us to the exponentiation of the 1-loop result did
not depend on the choice of a specific observable. Thus we expect that
the same resummation would also be applicable to other inclusive
quantities; the spectrum of fluctuations superposed on the Glasma
fields is universal.

The essential ingredient in eq.~(\ref{eq:master-formula}) is the small
fluctuations correlator $\Gamma_2(\u,\v)$, defined in
eq.~(\ref{eq:small-fluctuation-prop}), which should be computed with
the two endpoints on the initial time surface $\Sigma$, with the
Glasma field in the background. A first attempt to compute this object
is given in~\cite{FukusGM1}. However, the expression obtained there
was incomplete because it corresponds to an approximate expression for
the free propagator in $A^\tau=0$ gauge, with the only dependence on
the background field coming from Gauss's law. We will compute here (in
$A^\tau=0$ gauge) the spectrum of fluctuations both in the free case
and in the presence of a classical background field. We will show
later that the latter has a non-trivial dependence on the classical
fields in the Glasma at $\Sigma$.

\subsection{Gauge invariance of the spectrum of fluctuations}

The left hand side of the expression in eq.~(\ref{eq:master-formula})
should be gauge invariant because the energy-momentum tensor is a
physical quantity.  It should therefore be invariant under a gauge
transformation of the classical Glasma field,
\begin{equation}
{\cal A}
\longrightarrow \Omega^\dagger{\cal A}^\mu\Omega
+\frac{i}{g}\Omega^\dagger \partial^\mu \Omega\; .
\label{eq:GT-background}
\end{equation}
This invariance is also true for its leading order counterpart on the
right hand side of the expression, when we apply the same gauge
transformation to the total field ${\cal A}+\alpha$,
\begin{equation}
{\cal A} + \alpha
\longrightarrow \Omega^\dagger \left({\cal A}^\mu+\alpha^\mu\right)\Omega
+\frac{i}{g}\Omega^\dagger \partial^\mu \Omega\; .
\label{eq:GT-background1}
\end{equation}
Thus for the expression in eq.~(\ref{eq:master-formula}) to be
manifestly gauge invariant, it is sufficient if the initial spectrum
of fluctuations $F_0$ is invariant under the transformation
\begin{equation}
 \alpha
\longrightarrow \Omega^\dagger \alpha\Omega\; .
\label{eq:GT-background2}
\end{equation}
If we decompose $\alpha$ on the basis of the generators of $SU(3)$,
$\alpha\equiv \alpha_a t^a$, the identity
\begin{equation}
{\bar \Omega}_{ab} t^b = \Omega\; t^a \Omega^\dagger\; ,
\end{equation}
(where ${\bar \Omega}_{ab}$ here is an adjoint SU(3) matrix) gives
the equivalent transformation for the components $\alpha_a$ to be 
\begin{equation}
\alpha_a \to {\bar \Omega}_{ab} a^b\equiv \tilde\alpha_a\; .
\end{equation}
For the argument of the exponential in $F_0$ to be invariant under
this transformation,
\begin{equation}
\alpha_a(\u) \,\Gamma_{2,ab}^{-1}(\u,\v)\, \alpha_b(\v) \longrightarrow 
{\tilde\alpha}_a(\u)\,{\widetilde\Gamma}_{2,ab}^{-1}(\u,\v)\,
{\tilde\alpha}_b(\v) \; ,
\end{equation}
the inverse small fluctuations correlator in the Glasma (which is an
8$\times$8 adjoint matrix in SU(3)) must satisfy
\begin{equation}
{\widetilde\Gamma}_2^{-1}(\u,\v) = {\bar\Omega}(\u) \,\Gamma_2^{-1}(\u,\v)\,
{\bar\Omega}^\dagger(\v) \; .
\label{eq:gauge-invariance}
\end{equation}
It is clear from the structure of the expression in
eq.~(\ref{eq:small-fluctuation-prop}) that this property will be
satisfied. 

\subsection{Renormalization of ultraviolet divergences in the Glasma}

It is also important to address the potential ultraviolet divergences
in eq.~(\ref{eq:master-formula}). The leading order energy-momentum
tensor in the Glasma is ultraviolet finite. However, because one is
resumming quantum fluctuations in eq.~(\ref{eq:master-formula}), the
energy-momentum tensor should receive a contribution from the
(infinite) zero point energy. One can regularize ultraviolet
divergences by introducing a cutoff $\Lambda$ corresponding to the
largest momentum mode\footnote{In practical implementations of this
  resummation, space is discretized on a lattice, and thus the UV
  cutoff is the inverse of the lattice spacing.} of the fluctuation
$\alpha$. Since the energy-momentum tensor has canonical dimension
four, we expect that its dependence on this cutoff can be organized as
\begin{equation}
T^{\mu\nu}_{\rm resummed}(x)
=
c_1 \Lambda^4 +
c_2 \Lambda^2 +
c_3\; ,
\label{eq:Tmunu-div}
\end{equation}
where $c_{1,2,3}$ are finite quantities. It is easy to renormalize the
energy-momentum tensor by subtraction if one can prove that the
divergences are truly a property of the vacuum and do not depend on
the background classical field ${\cal A}$ in the Glasma. The
coefficient $c_1$ is dimensionless -- it is therefore a pure number,
that cannot depend on the background field. The case of $c_2$ is
trickier. Indeed, its canonical dimension 2 allows a priori a
dependence on the background field. However, we know that the left
hand side in eq.~(\ref{eq:Tmunu-div}) is invariant under gauge
transformations of the background field; we therefore must conclude
that the coefficient $c_2$ must be a gauge invariant, local (because
the left hand side is a local quantity), dimension 2 quantity. There
is no such quantity in Yang-Mills theory, which suggests that
$c_2=0$. Thus, on the basis of gauge symmetry and locality, we expect
that the only ultraviolet divergence in our expression for the
resummed energy-momentum tensor is a quartic divergence, with a
coefficient that does not depend on the background field. It can be
computed in principle once and for all in the absence of the
background field and subtracted from $T_{\rm resummed}^{\mu\nu}$ to
give an ultraviolet finite result for this quantity.

\section{Orthonormal basis of small fluctuations}
\label{sec:spectrum1}
\subsection{Introduction}
We noted in eq.~(\ref{eq:master-formula}) of the previous section
that the small fluctuations 2-point correlator $\Gamma_2(\u,\v)$ is
the key ingredient in resumming the contributions of leading
instabilities at all loop orders to the energy-momentum tensor.

In this section and the following one, we will work in the {\sl
  Fock-Schwinger} gauge $A^\tau=0$. Albeit at first sight a natural
gauge for describing hadron-hadron collisions (because it is an
interpolation between two light-cone gauges in the forward
light-cone), the {\sl Fock-Schwinger} gauge is not frequently used in
the literature. This is because even expressions for the free
correlator are complicated in this gauge. Our motivation here is
specific to the nature of the CGC description of heavy ion
collisions. The initial conditions for the evolution of classical
gauge fields in the forward light cone, in this gauge, are simply
expressed~\cite{KovneMW1} in terms of the classical fields in the
nuclei before the collision. This is an important criterion because
our resummed result in eq.~(\ref{eq:master-formula}) for the
energy-momentum tensor is expressed in terms of solutions of classical
Yang-Mills equations with Gaussian distributed initial
conditions. Further, numerical computations are unavoidable because
one is in a strong field regime where perturbative computations are
invalid; thus analytically cumbersome expressions are not a deterrent
if efficient numerical algorithms are feasible.

Turning to the computation of the small fluctuations correlator,
\begin{equation}
\Gamma_2 (\u,\v) = \sum_{\lambda,a}\int\frac{d^3\k}{(2\pi)^3 2k}
\;a_{+\k\lambda a}(\u) a_{-\k\lambda a}(\v)\; ,
\label{eq:G2-def}
\end{equation}
as noted previously in the discussion after eq.~(\ref{eq:shifts}), the
fluctuation fields $a_{\pm\k\lambda a}^\mu$ are plane wave fields at
$x^0=-\infty$ that have been evolved in time by interacting with the
classical background field ${\cal A}$. Before going further, let us
state two properties of this correlator that are true when the two
points $\u$ and $\v$ lie on the same Cauchy surface\footnote{For these
  properties to hold true, it is important that the two points are not
  separated by a time-like interval, which {\it ipso facto} is
  guaranteed if they belong to the same locally space-like surface.},
\begin{itemize}
\item[{\bf i.}] $\Gamma_2$ is symmetric: $\Gamma_2 (\u,\v)=\Gamma_2 (\v,\u)$,
\item[{\bf ii.}] $\Gamma_2$ is real valued.
\end{itemize}
Note that for the correlator as defined, the two properties are equivalent since
$a_{-\k\lambda a}=(a_{+\k\lambda a})^*$.  These properties are
crucial since $\Gamma_2$ is the variance of the fluctuations in our
resummation formula, eq.~(\ref{eq:master-formula}).

From the definition given in eq.~(\ref{eq:G2-def}), one might presume
that the calculation of $\Gamma_2$ requires one to follow the entire
evolution from $x^0=-\infty$ to $\tau=0^+$.  We will demonstrate in
section~\ref{sec:inner-product} that this is not the case; to compute
$\Gamma_2$, it is sufficient to construct an orthonormal basis of
fluctuation fields about classical fields at small proper times
$\tau=0^+$.

\subsection{Curvilinear coordinates}
The fluctuations $a_{\pm\k\lambda a}^\mu$ start as plane waves in the
remote past and evolve about the Glasma classical field ${\cal A}$
according to the equation of motion (\ref{eq:eom-fluct}). This
expression of the equation of motion, as well as the Yang-Mills
equation (\ref{eq:YM}), implicitly assume a Cartesian system of
co-ordinates. However, the most natural co-ordinate system in the
treatment of the post-collision evolution is the
$(\tau,\eta,\x_\perp)$ system, where
\begin{equation}
\tau \equiv \sqrt{t^2-z^2}\quad,\qquad 
\eta\equiv \frac{1}{2}\ln\left(\frac{t+z}{t-z}\right)\; ,
\end{equation}
and $\x_\perp$ collectively denotes the two co-ordinates perpendicular
to the collision axis. In these co-ordinates, the metric tensor
\begin{equation}
g_{\mu\nu}=\mbox{diag}\left(1,-1,-1,-\tau^2\right)
\end{equation}
has a proper time dependent determinant, $\sqrt{-g}=\tau$. This
implies small changes to the equations of motion. The classical
Yang-Mills equations (\ref{eq:YM}) become\footnote{Field equations
  can be generalized to an arbitrary system of coordinates by trading
  ordinary derivatives $\partial_\mu$ for covariant derivatives
  $\nabla_\mu$ (here, {\sl covariant} refers to coordinate
  transformations, not $SU(3)$ gauge transformations) that involve
  Christoffel symbols. For the Yang-Mills equations, one should thus
  write
\begin{equation*}
(\nabla_\mu -igA_\mu){\cal F}^{\mu\nu}=J^\nu\quad,\qquad
{\cal F}^{\mu\nu}\equiv \nabla^\mu A^\nu-\nabla^\nu A^\mu -ig[A^\mu,A^\nu]\; .
\end{equation*}
However, it turns out that for an {\sl antisymmetric} tensor such as
${\cal F}^{\mu\nu}$ one has also~(see \cite{Weinb3}, chapter 5)
\begin{equation*}
\nabla_\mu {\cal F}^{\mu\nu} = \frac{1}{\sqrt{-g}}
{\cal \partial}_\mu\left(\sqrt{-g}\,{\cal F}^{\mu\nu}\right)\; .
\end{equation*}
Moreover, one can show that $\nabla_\mu A_\nu-\nabla_\nu
A_\mu=\partial_\mu A_\nu-\partial_\nu A_\mu$, so that the flat-space
expression of ${\cal F}_{\mu\nu}$ can still be used (provided the two
indices are downstairs). In other words, the usual formulas for the
field strength should be used only for lower indices, and the metric
tensor should be used to raise indices if necessary. For instance, in
$A^\tau=A_\tau=0$ gauge, one has ${\cal F}_{\tau\eta}=\partial_\tau
A_\eta$ and ${\cal F}^{\tau\eta}=g^{\tau\tau}g^{\eta\eta}{\cal
  F}_{\tau\eta}=-\tau^{-2}\partial_\tau A_\eta$, but $\partial^\tau
A^\eta=\partial^\tau (-\tau^{-2}A_\eta)$ {\sl is not equal} to ${\cal
  F}^{\tau\eta}$. }
\begin{equation}
\frac{1}{\sqrt{-g}}
{\cal D}_\alpha\left(\sqrt{-g}\,g^{\alpha\beta}g^{\nu\mu}{\cal F}_{\beta\mu}\right)
=\delta^{\nu+}\rho_1+\delta^{\nu-}\rho_2\; ,
\label{eq:YM-curv}
\end{equation}
and the small fluctuation eqs.~(\ref{eq:eom-fluct}) become
\begin{equation}
\frac{1}{\sqrt{-g}}{\cal D}_\alpha
\left(\sqrt{-g}g^{\alpha\beta}g^{\nu\mu}\left({\cal D}_\beta a_\mu-{\cal D}_\mu a_\beta\right)\right)
-ig\,g^{\alpha\beta}g^{\nu\mu}
{\cal F}_{\mu\beta} \; a_\alpha=0\; .
\label{eq:eom-fluct-curv}
\end{equation}
In Fock-Schwinger gauge $A^\tau=0$ and $(\tau,\eta,\x_\perp)$ co-ordinates, these equations can be written out more explicitly to read,
\begin{eqnarray}
{\cal D}_\eta \partial_\tau a_\eta+\tau^2 {\cal D}_i \partial_\tau a_i -\frac{1}{2}\partial_\tau {\cal D}_\eta a_\eta -\frac{1}{2}\tau^2 \partial_\tau {\cal D}_i a_i&=&0\nonumber\\
\left(\partial_\tau \tau^{-1}\partial_\tau-\tau^{-1}\mathcal{P}_{ii}\right)a_\eta+\tau^{-1}\mathcal{P}_{i\eta} a_i&=&0 \nonumber\\
\left(\partial_\tau \tau \partial_\tau-\tau^{-1}\mathcal{P}_{\eta\eta}-\tau\mathcal{P}_{ii}\right)a_x+\tau^{-1}\mathcal{P}_{\eta x} a_\eta+\tau \mathcal{P}_{ix} a_i&=&0 \nonumber\\
\left(\partial_\tau \tau \partial_\tau-\tau^{-1}\mathcal{P}_{\eta\eta}-\tau\mathcal{P}_{ii}\right)a_y+\tau^{-1}\mathcal{P}_{\eta y} a_\eta+\tau\mathcal{P}_{iy} a_i&=&0\; ,
\label{eq:linqcd} 
\end{eqnarray}
where we have introduced the shorthand
${\mathcal{P}}$, which is defined as 
\begin{equation}
{\mathcal{P}}_{_{IJ}}^{ab} a_{_K}^b
\equiv
{\cal D}_{_I}^{ac} {\cal D}_{_J}^{cb} a_{_K}^b
+
g f^{acb} {\cal F}_{_{IJ}}^c a_{_K}^b \; .
\label{eq:projector}
\end{equation}
Here the indices $I,J,K$ denote $x,y$ and $\eta$, while we use Latin
indices $i,j$ to designate the two transverse coordinates $x,y$.  For
example,
$\partial_i^2=\partial_x^2+\partial_y^2=\partial_\perp^2$. Note that
the first equation in eq.~(\ref{eq:linqcd}) is Gauss' law, which can
also be written as
\begin{equation}
 {\cal D}_\eta \partial_\tau a_\eta
+
\tau^2 {\cal D}_i \partial_\tau a_i 
+ ig
\left(\partial_\tau {\cal A}_\eta\right) a_\eta
+
\tau^2 \left(\partial_\tau {\cal A}_i\right) a_i=0 \; .
\end{equation}

Even though we exclusively work in $\tau$--$\eta$ coordinates, we
shall at times make use of light-cone coordinates defined as
\begin{equation}
x^{\pm}\equiv\frac{t\pm z}{\sqrt{2}}\; ,
\end{equation}
with the metric tensor now having the form
\begin{equation}
g_{+-}=g_{-+}=1\spc\spc,\spc\spc g_{xx}=g_{yy}=-1 \; .
\end{equation}
The transformation from light-cone to $\tau,\eta$ coordinates is given by
\begin{equation}
x^{\pm}=\frac{\tau\, e^{\pm\eta}}{\sqrt{2}}\; ,
\end{equation}
while the Fock-Schwinger gauge condition, expressed in light-cone coordinates, is 
\begin{equation}
A^\tau=A_\tau=\frac{1}{\tau}\left(x^+A^-+x^-A^+\right)=0 \; .
\end{equation}

\subsection{Inner product for small field fluctuations}
\label{sec:inner-product}
Since the equations of motion (\ref{eq:eom-fluct-curv}) of the small
fluctuations are linear, the set of its solutions is a vector space,
and it is sufficient to know a basis of this space in order to be able
to construct any solution.  For a real background field such as the
classical field ${\cal A}^\mu$, the evolution in time of the small
fluctuations is unitary\footnote{In event of confusion from the
  apparent structure of the last term of
  eq.~(\ref{eq:eom-fluct-curv}), note that the components of the
  adjoint generators are purely imaginary, and therefore the function
  that multiplies $a^\mu$ in this term is real.}. Therefore, there
should be an inner product between pairs of solutions of
eq.~(\ref{eq:eom-fluct-curv}) that remains invariant during the
evolution of these solutions. To construct this inner product, rewrite
eq.~(\ref{eq:eom-fluct-curv}) as
\begin{equation}
{\cal O}^{\nu\mu} a_\mu =0\; ,
\label{eq:fluct2}
\end{equation}
with
\begin{equation}
{\cal O}^{\nu\mu}
\equiv
D_\alpha \sqrt{-g}
\Big(
g^{\nu\mu}g^{\alpha\beta}
-g^{\nu\beta}g^{\mu\alpha}
\Big) D_\beta 
-ig \sqrt{-g} g^{\nu\alpha}g^{\mu\beta}{\cal F}_{\alpha\beta}
\; .
\label{eq:O-def}
\end{equation}
Consider now two solutions $a_\mu$ and $b_\mu$ of
eq.~(\ref{eq:fluct2}), and start from the identity
\begin{equation}
0=\int_\Omega d^4x\; a^*_\nu(x)\, 
\Big[
\stackrel{\longrightarrow}{{\cal O}^{\nu\mu}}
-
\stackrel{\longleftarrow}{{\cal O}^{\nu\mu*}}
\Big]\,b_\mu(x)\; ,
\label{eq:zero}
\end{equation}
where $\Omega$ is some domain of space-time. This identity is a trivial
consequence of the equations of motion for $a^*$ and $b$. A remarkable
property of the integrand in the right hand side is that it is a total
derivative\footnote{For this property to be true, it is crucial that
  the last term in eq.~(\ref{eq:O-def}) is real and one should
  properly take the complex conjugate of the covariant derivatives
  when they act on the left. This property is in fact closely related
  to the operator ${\cal O}^{\nu\mu}$ being Hermitean; the evolution of the fluctuations is unitary.},
\begin{eqnarray}
a^*_\nu(x)\, 
\Big[
\stackrel{\longrightarrow}{{\cal O}^{\nu\mu}}
-
\stackrel{\longleftarrow}{{\cal O}^{\nu\mu*}}
\Big]\,b_\mu(x)
=
\partial_\alpha\Big[\sqrt{-g}\Big(
g^{\nu\mu}g^{\alpha\beta}
-\frac{1}{2}g^{\nu\beta}g^{\mu\alpha}
-\frac{1}{2}g^{\nu\alpha}g^{\mu\beta}
\Big)
\Big(
a_\nu^*(x)
\stackrel{\leftrightarrow}{\cal D}_\beta
b_\mu(x)\Big)
\Big]
\; .
\end{eqnarray}
Therefore, one can use Stokes theorem,
\begin{equation}
\int_\Omega d^4x\;\partial_\alpha F^\alpha
=
\int_{\partial\Omega}d^3{\bs S}_u\;n_\alpha F^\alpha
\end{equation}
where $d^3{\bs S}_\u$ is the measure on the boundary $\partial\Omega$, and
$n_\alpha$ is a normal vector to the boundary, oriented outwards. Let us
assume that the boundary $\partial\Omega$ is made of two locally
space-like surfaces $\Sigma_1$ and $\Sigma_2$, and a third boundary
located at infinity in the spatial directions on which all the fields
are vanishing. Then eq.~(\ref{eq:zero}) is equivalent to
\begin{eqnarray}
&&
\int_{\Sigma_1}d^3{\bs S}_\u\;\sqrt{-g}\Big(
g^{\nu\mu}g^{\alpha\beta}
-\frac{1}{2}g^{\nu\beta}g^{\mu\alpha}
-\frac{1}{2}g^{\nu\alpha}g^{\mu\beta}
\Big)
n_\alpha \Big(a^*_\nu(u)
\stackrel{\leftrightarrow}{\cal D}_\beta
b_\mu(u)\Big)
\nonumber\\
&=&
\int_{\Sigma_2}d^3{\bs S}_\u\;\sqrt{-g}\Big(
g^{\nu\mu}g^{\alpha\beta}
-\frac{1}{2}g^{\nu\beta}g^{\mu\alpha}
-\frac{1}{2}g^{\nu\alpha}g^{\mu\beta}
\Big)
n_\alpha \Big(a^*_\nu(u)
\stackrel{\leftrightarrow}{\cal D}_\beta
b_\mu(u)\big)\; .
\end{eqnarray}
We have thus proved, most generally, that an inner product defined as 
\begin{equation}
\big(a\big|b\big)
\equiv i
\int_{\Sigma}d^3{\bs S}_\u\;\sqrt{-g}\Big(
g^{\nu\mu}g^{\alpha\beta}
-\frac{1}{2}g^{\nu\beta}g^{\mu\alpha}
-\frac{1}{2}g^{\nu\alpha}g^{\mu\beta}
\Big)
n_\alpha\Big(a^*_\nu(u)
\stackrel{\leftrightarrow}{\cal D}_\beta
b_\mu(u)\Big)\; ,
\label{eq:inner-product}
\end{equation}
is independent of the Cauchy surface $\Sigma$ used to define it,
provided $a_\mu$ and $b_\mu$ obey the equation of motion of small
fluctuations. Note that we have added a factor $i$ to its definition
to ensure  that it is Hermitean,
\begin{eqnarray}
\big(a\big|b\big)^*&=& \big(b\big|a\big)\; ,\nonumber\\
\big(a^*\big|b^*\big)&=&-\big(b\big|a\big)=-\big(a\big|b\big)^*\; .
\label{eq:inner-prod-properties}
\end{eqnarray}
In the special case where $\Sigma$ is a surface of constant $\tau$ and
we work in the Fock-Schwinger gauge $A^\tau=0$, we have
$n\cdot{\cal D}=\partial_\tau$, and $n\cdot a=0$, $n\cdot
b=0$. Therefore the inner product simplifies into
\begin{equation}
\big(a\big|b\big)
\equiv i\!\!\int\limits_{\tau=\mbox{const}}\!\!
d^3{\bs S}_\u\;\sqrt{-g}\;g^{\nu\mu}
\Big(a^*_\nu(u)\,\stackrel{\leftrightarrow}{\partial}_\tau\, b_\mu(u)\Big)\; .
\label{eq:inner-product-1}
\end{equation}

Now let us evaluate the inner product for pairs of field fluctuations
taken from the set of the $a_{\pm\k\lambda a}$. Since the inner
product does not depend on the chosen time surface and since we know
these fields at $x^0\to -\infty$ (because they are defined via their
initial condition in the remote past), we can evaluate the inner product by using plane wave initial conditions for these fluctuation fields. This gives
\begin{eqnarray}
\big(a_{+\k\lambda a}\big|a_{-\l\rho b}\big) &=& 0
\nonumber\\
\big(a_{+\k\lambda a}\big|a_{+\l\rho b}\big) &=& \delta_{\lambda\rho} \delta_{ab} (2\pi)^3 2k \delta(\k-\l)
\nonumber\\
\big(a_{-\k\lambda a}\big|a_{-\l\rho b}\big) &=&-\delta_{\lambda\rho} \delta_{ab} (2\pi)^3 2k \delta(\k-\l)
\; .
\label{eq:normalization}
\end{eqnarray}
Thus this particular basis of the space of solutions of
eq.~(\ref{eq:eom-fluct}) is orthonormal with respect to the invariant
inner product defined in eq.~(\ref{eq:inner-product}). Note also that
the $a_{+\k\lambda a}$'s represent only one {\sl half of the basis} of
the vector space of solutions of eq.~(\ref{eq:eom-fluct-curv})
--namely the solutions that have a positive frequency in the remote
past. The other half is simply obtained by complex conjugation. It
easy to check that any unitary transformation of the positive energy
solutions (and a concomitant change to the negative energy ones, that
are their complex conjugates) transforms an orthonormal basis into
another orthonormal basis, and leaves the formula
eq.~(\ref{eq:G2-def}) unchanged.  This remark is useful because it
leaves us the freedom to label the elements of the basis by other
quantities than the Cartesian $3$-momentum. This will be true in our
specific case where we are interested in a basis in a curvilinear
co-ordinate system.

\subsubsection{Normalization of the fields and choice of basis}

It is important to note that the prefactor in front of the $\delta$
functions in eq.~(\ref{eq:normalization}) exactly cancels the factors
that are included in the integration measure in eq.~(\ref{eq:G2-def}),
namely one has
\begin{equation}
\int\frac{d^3\k}{(2\pi)^3 2k}\;\big(a_{+\k\lambda a}\big|a_{+\l\rho b}\big)=1\; .
\end{equation}
This remark in fact defines uniquely how the inner product of the
basis elements should be normalized given a generic choice for the
integration measure in eq.~(\ref{eq:G2-def}); in particular, this rule
will be come in handy later when we use other labels than the usual
3-momentum to index the elements of the basis. Moreover, this makes
clear that eq.~(\ref{eq:G2-def}) is just one particular representation
of the correlator $\Gamma_2$; there exists such a representation for
any orthonormal basis of the space of solutions of
eq.~(\ref{eq:eom-fluct-curv}), as we shall explain now. Thanks to the
above inner product, one can spell out a general procedure\footnote{In
  this light, eq.~(\ref{eq:G2-def}) which represents $\Gamma_2$ in
  terms of the $a_{\pm\k\lambda a}$'s, exploits one possible method of
  constructing such an orthonormal basis. In this case, one starts at
  $x^0=-\infty$ with the plane waves, that are known to form an
  orthonormal basis, and evolves them forward to the time of
  interest. The time invariance of the inner product then guarantees
  us that we get an orthonormal basis on the forward light-cone.}  for
constructing the correlator $\Gamma_2$:
\begin{itemize}
\item[{\bf i.}] Find a complete set of independent {\sl positive
  energy} solutions $a_{_K}$ of eq.~(\ref{eq:eom-fluct-curv}), where
  $K$ denotes collectively (usually a mix of continuous and discrete
  labels) all the labels necessary to index these solutions.
\item[{\bf ii.}] This set of solutions should obey the 
  orthogonality condition,
  \begin{equation}
    \big(a_{_K}\big|a_{_{K^\prime}}\big)=N_{_K}\,\delta_{_{KK^\prime}}
    \label{eq:ortho}
  \end{equation}
  with $N_{_K}$ real and positive definite\footnote{This means that
    the solutions $a_{_K}$ will in general be complex solutions.},
\item[{\bf iii.}] The correlator $\Gamma_2$ is then given by
  \begin{equation}
    \Gamma_2(\u,\v)=\int d\mu_{_K}\;a_{_K}(\u)a_{_K}^*(\v)\; ,
    \label{eq:G2-generic}
  \end{equation}
  where the measure $d\mu_{_K}$ (a mix of integrals and discrete sums)
  is such that
  \begin{equation}
    \int d\mu_{_K}\;N_{_K}\,\delta_{_{KK^\prime}}=1\; .
    \label{eq:measure}
  \end{equation}
\end{itemize}
It is clear from eqs.~(\ref{eq:ortho}) and (\ref{eq:measure}) that the
$\Gamma_2$ given by eq.~(\ref{eq:G2-generic}) is independent of how we
normalize the solutions (i.e. on the constants $N_{_K}$), provided we
choose the integration measure accordingly.  Moreover, we only need to
know the form of the solutions at the time of interest, and we can
avoid the complication of evolving the plane waves from the past
through the forward light-cone. This is particularly helpful when one
uses $\tau,\eta$ coordinates, because this system of coordinates has a
singularity at $\tau=0$.

A further simplification is possible because in practice we won't need
to use directly eq.~(\ref{eq:G2-generic}) for $\Gamma_2$. Indeed, an
ensemble of real-valued field fluctuations $a^\mu$ that have a 2-point
equal-time correlation given by $\Gamma_2$ can be generated by the
following formula,
\begin{equation}
a^\mu(x)=\int d\mu_{_K}\;\Big[c_{_K}\,a_{_K}^\mu(x)+c_{_K}^*\,a_{_K}^{\mu*}(x)\Big]\; ,
\label{eq:random1}
\end{equation}
where the coefficients $c_{_K}$ are random Gaussian-distributed
complex numbers whose variance is given by
\begin{eqnarray}
\left<c_{_K}c_{_{K^\prime}}^*\right>&=&\frac{N_{_K}}{2} \delta_{_{KK^\prime}}
\nonumber\\
\left<c_{_K}c_{_{K^\prime}}\right>&=&\left<c_{_K}^*c_{_{K^\prime}}^*\right>=0 \; .
\label{eq:random2}
\end{eqnarray}
This method of generating the field fluctuations offers the advantage
that it does not require that one diagonalizes the correlation
function $\Gamma_2$.

\subsection{Free fluctuations in $A^\tau=0$ gauge}
Let us start by calculating the fluctuation correlator in $A^\tau=0$
gauge and on the surface $\tau=0^+$ in the {\sl free case}, namely, in
the absence of the classical background field. Given the complications
introduced by the choice of gauge and the system of curvilinear
coordinates, this is a useful exercise to pursue before attacking the
more general case of the Glasma background field.  In this situation,
eqs.~(\ref{eq:linqcd}) simplify into
\begin{eqnarray}
\partial_\eta \partial_\tau a_\eta+\tau^2 \partial_i \partial_\tau a_i &=&0\nonumber\\
\left(\partial_\tau \tau^{-1}\partial_\tau-\tau^{-1}\partial_\perp^2\right)a_\eta+\tau^{-1}\partial_i\partial_\eta a_i&=&0 \nonumber\\
\left(\partial_\tau \tau \partial_\tau-\tau^{-1}\partial_\eta^2
-\tau\partial_\perp^2\right)a_x+\tau^{-1}\partial_\eta\partial_x a_\eta
+\tau \partial_i\partial_x a_i&=&0 \nonumber\\
\left(\partial_\tau \tau \partial_\tau
-\tau^{-1}\partial_\eta^2
-\tau\partial_\perp^2\right)a_y
+\tau^{-1}\partial_\eta\partial_y a_\eta+\tau\partial_i\partial_y a_i&=&0\; .
\label{eq:linqcd-free} 
\end{eqnarray}

\subsubsection{Residual gauge freedom}
A general solution $a^\mu$ to eq.~(\ref{eq:linqcd-free}) has a priori
four components.  However, a massless vector field has only two
physical degrees of freedom.  One of the seemingly independent
components of the vector field is removed by the gauge condition
$a_\tau=a^\tau=0$ (this is already implemented in
eqs.~(\ref{eq:linqcd-free})).  However, even after imposing this
condition, there is a residual gauge symmetry in the equations of
motion, namely, these are invariant under $\tau$ independent gauge
transformations
\begin{equation}
a^\mu \to a^\mu + \partial^\mu \Lambda(\eta,\x_\perp)\; ,
\end{equation}
where $\Lambda$ is an arbitrary $\tau$ independent function.  As a
consequence of this residual gauge freedom, the three remaining
components of $a^\mu$ are not all physical degrees of freedom.

In order to find the two physical solutions, we begin by finding the
unphysical solution.  This solution must be a {\sl pure gauge}, which
here means it is a $\tau$ independent total derivative.  As we will
see, the unphysical solution is not a dynamical variable but is
completely constrained by the initial and boundary conditions.  After
finding the most general $\tau$ independent solution to the equations
of motion, the two physical solutions can be determined relatively easily.  Their
form will be narrowed down by requiring that the three solutions are
mutually orthonormal, and then the residual gauge freedom will be fixed by
imposing the equations of motion.

In the $(\tau,\eta,\x_\perp)$ system of coordinates, a convenient set
of labels for the solutions of eq.~(\ref{eq:linqcd-free}) is
$\nu,\k_\perp,\lambda,a$, where $\nu$ is the Fourier conjugate of the
space-time rapidity $\eta$ (as used previously, $\lambda$ denotes the
polarization and $a$ the color). We choose $\lambda=1,2$ to denote the
physical solutions, and $\lambda=3$ to be the unphysical one. Since in
this section we are in the vacuum, all the colors $a$ obey exactly the
same equations and we can drop this index to keep the notations
lighter. With this set of variables, eqs.~(\ref{eq:ortho}),
(\ref{eq:G2-generic}) and (\ref{eq:measure}) take the form
\begin{eqnarray}
  \Gamma_2(\bfx,\bfx^\prime)&=&
  \sum_{\lambda=1,2}\int \frac{d^2\bfk_\perp d\nu}{(2\pi)^3}\;
  a_{\bfk\lambda}(\tau=0^+,\bfx) a_{\bfk\lambda}^*(\tau=0^+,\bfx^\prime)\; ,
  \nonumber\\
  \big(a_{\bfk\lambda}\big|a_{\bfk^\prime\lambda^\prime}\big)&=&(2\pi)^3
\delta_{\lambda \lambda^\prime}
\underbrace{\delta(\nu-\nu^\prime)\delta(\k_\perp - \k_\perp^\prime)}_{\delta(\bfk-\bfk^\prime)}\; ,
\label{eq:norm}
\end{eqnarray}
where we use the shorthands $\bfk\equiv(\nu,\k_\perp)$ and $\bfx\equiv
(\eta,\x_\perp)$.

\subsubsection{Vacuum solutions}
In a linear system of coordinates, the $a_{\bfk\lambda}^\mu(x)$
introduced above would have the simple following parametrization,
\begin{equation}
a^\mu_{\bfk\lambda}(x)
=
e^{i\bfk\cdot\bfx}
e^{-i\omega_\k x^0}
\;
\varepsilon^\mu_{\bfk\lambda}\; ,
\label{eq:pos-energy-wave}
\end{equation}
where $\varepsilon^\mu_{\bfk\lambda}$ is a constant polarization
vector.  Note that the minus sign in the exponential that gives the
time dependence is necessary in order to ensure that
$a_{\k\lambda}^\mu$ is a positive energy solution. But because we work
in a curvilinear co-ordinate system, the time dependence of the
solutions cannot be a simple exponential. Let us generalize the
previous expression by writing
\begin{equation}
a^\mu_{\bfk\lambda}(\tau,\bfx)
=
e^{i\bfk\cdot\bfx}
\;
\alpha^\mu_{\bfk\lambda}(\tau)\; ,
\label{eq:pol-curv}
\end{equation}
where we have combined the polarization vector and the time dependence
in a unique quantity that we denote
$\alpha^\mu_{\bfk\lambda}(\tau)$. (Since the equations of motion do
not have coefficients that depend explicitly on $\eta$ or $\x_\perp$,
it is clear that we can still look for solutions whose $\eta$ and
$\x_\perp$ dependence is of the form $\exp(i\bfk\cdot\bfx)$.) Here
again, we will have to make sure that the
$a^\mu_{\bfk\lambda}(\tau,\bfx)$ constructed in this way contains only
positive energy contributions.

Let us consider first the unphysical solution. This is a pure
gauge solution independent of $\tau$.  The most general $\tau$
independent solution is of the form
\begin{equation}
\alpha^\mu_{\bfk3}=\left(\begin{array}{c} \k_x \\ \k_y \\ \nu \end{array}\right)
\alpha_3(\nu,\k_\perp)\;,
\label{eq:unphysical}
\end{equation}
where $\alpha_3(\nu,\k_\perp)$ is an arbitrary function. The
inner product of the unphysical fluctuation $a^\mu_{\bfk3}$ with one
of the physical solutions ($a^\mu_{\bfk^\prime\lambda}$ with
  $\lambda=1,2$) is
\begin{eqnarray}
\big(a_{\bfk3}\big|a_{\bfk^\prime\lambda}\big)
&=&
 i\tau\alpha^{*}_3(\nu,\k_\perp) 
\int_{-\infty}^{+\infty}d\eta \int d^2\x_\perp\;e^{i(\bfk^\prime-\bfk)\cdot\bfx}\;
\partial_\tau
\left(
\k_x \alpha^x_{\bfk^\prime\lambda}
+
\k_y\alpha^y_{\bfk^\prime\lambda}
+
\nu\tau^{-2}  \alpha^\eta_{\bfk^\prime\lambda}
\right)
\nonumber\\
&=&
i\tau (2\pi)^3\delta(\bfk-\bfk^\prime)
\alpha^{*}_3(\nu,\k_\perp) \;
\partial_\tau
\left(
\k_x \alpha^x_{\bfk\lambda}
+
\k_y\alpha^y_{\bfk\lambda}
+
\nu\tau^{-2}  \alpha^\eta_{\bfk\lambda}
\right)
\; .
\label{eq:normunphys}
\end{eqnarray}
We can satisfy this orthogonality condition by choosing a second solution
of the form
\begin{equation}
\alpha^\mu_{\bfk1}(\tau)=
\left(\begin{array}{c} \k_y \\ -\k_x \\ 0 \end{array}\right)
\alpha_1(\tau,\nu,\k_\perp) \; .
\label{eq:a10-1}
\end{equation}
The functional form of $\alpha_1$ should be fixed by the equations of
motion.  Substituting the above expression into the equations of
motion (\ref{eq:linqcd-free}) yields a differential equation for
$\alpha_1$, whose general solution can be expressed in terms of the
Hankel functions $H^{(1)}_{i\nu}$ and $H^{(2)}_{i\nu}$,
\begin{equation}
\alpha_1(\tau,\nu,\k_\perp)
=
a_\bfk\, H^{(1)}_{i\nu}\left(k_\perp \tau\right)
+
b_\bfk\, H^{(2)}_{i\nu}\left(k_\perp \tau\right)\; .
\label{eq:alpha1}
\end{equation}
Recall here the integral representation of the Hankel functions,
\begin{eqnarray}
H^{(1)}_{i\nu}(x)&=&-\frac{i}{\pi}e^{+\pi\nu/2}\int_{-\infty}^{+\infty}
e^{ix\cosh t+i\nu t}dt\nonumber\\
H^{(2)}_{i\nu}(x)&=&+\frac{i}{\pi}e^{-\pi\nu/2}\int_{-\infty}^{+\infty}
e^{-ix\cosh t-i\nu t}dt \;,
\label{eq:Hankel1}
\end{eqnarray}
The convention set by eq.~(\ref{eq:pos-energy-wave}) implies that only $H^{(2)}_{i\nu}(k_\perp\tau)$ has the
appropriate frequency sign to be one the $a_{\bfk\lambda}$'s. We can therefore 
set $a_\bfk=0$ and keep only the second term in
eq.~(\ref{eq:alpha1}). The value of $b_\bfk$ can then be determined
by the orthogonality condition. For this, we need the identity
\begin{equation}
H_{i\nu}^{(2)*}(x)\stackrel{\leftrightarrow}{\partial}_x H_{i\nu}^{(2)}(x)
=
-\frac{4ie^{-\pi\nu}}{\pi x}\; ,
\end{equation}
from which we obtain
\begin{equation}
\big(a_{\bfk1}\big|a_{\bfk^\prime1}\big)
=
(2\pi)^3\delta(\bfk-\bfk^\prime) |b_\bfk|^2 \k_\perp^2 
\frac{4e^{-\pi\nu}}{\pi}\; .
\end{equation}
In order to get the same normalization as in eq.~(\ref{eq:norm}), we
obtain (up to an irrelevant phase)
\begin{equation}
b_\bfk = \frac{\sqrt{\pi}e^{\pi\nu/2}}{2\k_\perp}\; ,
\end{equation}
and the first physical solution reads
\begin{equation}
a_{\bfk1}^\mu(\bfx)
=
\frac{\sqrt{\pi}e^{\pi\nu/2}}{2\k_\perp}
\left(\begin{array}{c} \k_y \\ -\k_x \\ 0 \end{array}\right)
e^{i\bfk\cdot\bfx}\,
H_{i\nu}^{(2)}(\k_\perp\tau)\; .
\label{eq:zeroth-order-1-free}
\end{equation}

The second physical solution can be found by requiring that it be 
orthogonal to the two solutions we have so far.  This restricts it
to be of the form
\begin{equation}
\alpha_{\bfk2}^\mu
=
\left(\begin{array}{c} 
\nu \k_x\, \alpha_{2\perp}(\tau,\nu,\k_\perp)\\ 
\nu \k_y\, \alpha_{2\perp}(\tau,\nu,\k_\perp)\\ 
\;\;\;\;-\,\alpha_{2\eta}(\tau,\nu,\k_\perp)\  \end{array}\right)\; ,
\label{eq:alpha2}
\end{equation}
provided that $k_\perp^2\tau^2 \partial_\tau
\alpha_{2\perp}=\partial_\tau \alpha_{2\eta}$.  This last constraint
was derived by substituting the general form of the second solution
eq.~(\ref{eq:alpha2}) into the orthogonality condition
eq.~(\ref{eq:normunphys}) with the unphysical solution.  It turns out
that this is the same constraint needed for this solution to fulfill
Gauss's law.

Dynamical equations for the functions $\alpha_{2\perp}$ and
$\alpha_{2\eta}$ can be found by substituting eq.~(\ref{eq:alpha2})
into eqs.~(\ref{eq:linqcd-free}). One obtains, 
\begin{eqnarray}
\partial_\tau^3 \alpha_{2\eta} -\frac{1}{\tau}\partial_\tau^2 \alpha_{2\eta}+\left(\frac{\nu^2+1}{\tau^2}+\k_\perp^2\right)\partial_\tau \alpha_{2\eta}&=&0\;,\nonumber\\
\partial_\tau^3 \alpha_{2\perp} +\frac{3}{\tau}\partial_\tau^2 \alpha_{2\perp}+\left(\frac{\nu^2+1}{\tau^2}+\k_\perp^2\right)\partial_\tau \alpha_{2\perp}&=&0\; .
\label{eq:soln2de}
\end{eqnarray}
The positive energy solutions to the above third order differential
equations can be written as
\begin{eqnarray}
\alpha_{2\eta}&=&\mbox{const}\times\int^\tau\;d\tau^\prime \tau^\prime H^{(2)}_{i\nu}(\k_\perp\tau^\prime)\nonumber\\ 
\alpha_{2\perp}&=&\mbox{const}\times\int^\tau\; \frac{d\tau^\prime}{\tau^\prime}H^{(2)}_{i\nu}(\k_\perp\tau^\prime)\; .
\label{eq:soln2deint}
\end{eqnarray}
Note that the differential equations above imply that the functional
form of $\alpha_{2\perp}$ and $\alpha_{2\eta}$ are only determined up
to an arbitrary $\tau$ independent function.  This corresponds to a
residual gauge freedom in which we can always add to the second
physical solution eq.~(\ref{eq:alpha2}) a pure gauge solution having
the form of eq.~(\ref{eq:unphysical}).  This residual gauge freedom
can be removed by imposing an additional gauge fixing condition.  For
example, in simulations of classical Yang--Mills one typically imposes
transverse Coulomb gauge.

The integrals over the Hankel functions can be written in terms of
hypergeometric functions but their form is not very enlightening.  To streamline our notation, following Makhlin~\cite{Makhl3}, let
us define
\begin{equation}
R_{b,\alpha}^{(a)}(\k_\perp\tau)\equiv
\int^\tau dx\; x^{b} H_{\alpha}^{(a)}(\k_\perp x)\; .
\end{equation}
The properly normalized form for the second physical degree of freedom is
\begin{equation}
  a_{\bfk2}^\mu(\tau,\eta,\x_\perp)
  =\frac{\sqrt{\pi}e^{\pi\nu/2}}{2\k_\perp} \left(\begin{array}{c}  \,\nu\k_x R_{-1,i\nu}^{(2)}\left(\k_\perp\tau\right) \\ \nu\k_y R_{-1,i\nu}^{(2)}\left(\k_\perp \tau\right)\\ \;\;\;\;\;-R_{+1,i\nu}^{(2)}\left(\k_\perp\tau\right) \end{array}\right)\,e^{i\bfk\cdot\bfx}\; .
\label{eq:zeroth-order-2-free}
\end{equation}
For $\k_\perp\tau \ll 1$, we can make use of the series expansion of
the Hankel functions and rewrite the solution as
\begin{equation}
a_{\bfk2}^\mu(\bfx)
\approx
\frac{\sqrt{\pi}e^{\pi\nu/2}}{2\k_\perp}
\left(\begin{array}{c} \k_x \\ \k_y \\ -(\k_\perp\tau)^2/(\nu+2i) \end{array}\right)
e^{i\bfk\cdot\bfx}\;
H_{i\nu}^{(2)}(\k_\perp\tau)\; .
\label{eq:zeroth-order-2-free-asy}
\end{equation}
From these explicit solutions, we will construct in appendix
\ref{sec:gamma2} the correlation function $\Gamma_2$ for free fields
on the initial surface $\tau=0^+$. However, for the purposes of
generating a Gaussian ensemble of fluctuations with the proper
variance, the above results along with eqs.~(\ref{eq:random1}) and
(\ref{eq:random2}) are sufficient.

\section{Small fluctuations in the Glasma}
\label{sec:glasma}
After our extended discussion of the free fluctuations in $A^\tau=0$
gauge, we now turn to the derivation of the small fluctuations
spectrum in the Glasma.  We shall first write down the small
fluctuation equations of motion in the presence of the background
classical fields. The main difficulty here is that the Glasma
background fields are not known analytically at arbitrary proper
times. They are however known in closed form at $\tau = 0^+$ in terms
of the classical CGC fields before the collision. We will perform a
small time expansion, valid at very short proper times $\tau\ll
Q_s^{-1}$, of both the classical fields and the small fluctuation
fields and show that the fluctuations only depend on the classical
gauge fields immediately after the collision. As our final result, we
will obtain explicit expressions for an orthonormal basis of small
fluctuations that generalize eqs. (\ref{eq:zeroth-order-1-free}) and
(\ref{eq:zeroth-order-2-free}) to the case of a non-zero background field at
small proper times.

\subsection{Structure of the Glasma background field}
In the Fock--Schwinger gauge, the classical gauge field configurations
can be expressed as~\cite{KovneMW1,KrasnV3}
\begin{eqnarray}
{A}^i&=&\theta(-x^+)\theta(x^-)\alpha^i_{1}+\theta(x^+)\theta(-x^-)\alpha^i_{2} +\theta(x^+)\theta(x^-){\cal A}^i\nonumber\\
{A}^\eta&=& \theta(x^+)\theta(x^-){\cal A}^\eta
\label{eq:IC}
\end{eqnarray}
The fields $\alpha_{1,2}^i$ are the color fields of the two nuclei
before the collision, that take the form of transverse pure gauge
fields, while ${\cal A}^\mu$ denotes the classical fields after the collision.  

Since we are interested in the spectrum of fluctuations at $\tau=0^+$
we need only the behavior of the background fields shortly after the
collision.  The classical Glasma fields in the forward light cone can
be expanded, in all generality, at early times as
\begin{eqnarray}
{\cal A}_I &=& \sum_{n=0}^\infty {\cal A}_{(n)I} \tau^n
\label{eq:taylor}
\end{eqnarray}
The initial conditions for these background
fields at $\tau=0^+$ are obtained by matching the Yang-Mills equations
just below and just above the forward light-cone (to ensure a regular
behavior of the field equations). One
obtains~\cite{KovneMW1} for the fields and their
time derivatives,
\begin{eqnarray}
{\cal A}^{i}(\tau=0^+)&=&\alpha^i_1+\alpha^i_2 \nonumber\\
{\cal A}_{\eta}(\tau=0^+) &=& 0\nonumber\\
\mathcal{E}_{i}(\tau=0^+)&=&\tau \partial_\tau {\cal A}_i|_{\tau=0} = 0 \nonumber\\
\mathcal{E}^\eta(\tau=0^+)&=& \frac{1}{\tau} \partial_\tau {\cal A}_\eta|_{\tau=0}
= ig\left[ \alpha^i_{1},\alpha^i_{2}\right]\; .
\label{eq:class-IC}
\end{eqnarray}
As alluded to previously, explicit analytical solutions are known for the fields $\alpha_{1,2}^i$. 
The Taylor expansions of eqs.~(\ref{eq:taylor}) begin with
\begin{equation}
{\cal A}_i = {\cal A}_{i}(0^+)+{\cal O}(\tau)\quad,\qquad
{\cal A}_\eta = \frac{1}{2}{\cal E}^\eta(0^+)\tau^2+{\cal O}(\tau^3)\; .
\end{equation}

At this point it is useful to introduce some extra notation.  Based on the Taylor expansion in eq.~(\ref{eq:taylor}) of the classical
background field, it will be convenient to introduce an analogous
expansion of the covariant derivatives\footnote{Of course, the
  $D_\tau$ derivative does not need to be expanded, since in the
  Fock-Schwinger gauge one has $D_\tau\equiv \partial_\tau$.} and projectors
as defined in eq.~(\ref{eq:projector}),
\begin{eqnarray}
{\cal D}_\mu &=& \sum_n \tau^n {\cal D}_{(n)\mu}\;,\nonumber\\
{\cal P}_{\mu\nu}&=&\sum_n \tau^n {\cal P}_{(n)\mu\nu}\; .
\label{eq:DPexp}
\end{eqnarray}
Naturally, the coefficients ${\cal D}_{(n)\mu}$ and ${\cal
  P}_{(n)\mu\nu}$ can be written in terms of the Taylor coefficients
of the classical background field, $\mathcal{A}_{(n)I}$.  For example, if we perform the Taylor expansion of the covariant derivative, ${\cal D}_{(n)\mu}$, the leading coefficients can be expressed in terms of the Taylor coefficients of the classical background field,
\begin{eqnarray}
{\cal D}_{(0)i}^{ab}=\delta^{ab}\partial_i-ig\mathcal{A}_{(0)i}^{ab}\;\;\; , \;\;\; {\cal D}_{(0)\eta}^{ab}=\delta^{ab}\partial_\eta\;\;\; , \;\;\; {\cal D}_{(2)\eta}^{ab}=-ig\mathcal{A}_{(2)\eta}^{ab}\; . 
\label{eq:Dexp}
\end{eqnarray}
In writing down the above expressions we have used the fact that the transverse components of the background field have non--vanishing zero order Taylor coefficients in contrast to the longitudinal component of the background field whose leading behavior starts with $\tau^2$.  Similarly we can express the Taylor coefficients of ${\cal  P}_{(n)\mu\nu}$ in terms of the Taylor coefficients of the classical background field.  The terms which will needed later on in our discussion include,
\begin{eqnarray}
{\cal P}_{(0)\eta i}^{ab}&=&{\cal P}_{(0)i\eta}^{ab}={\cal D}_{(0)i}^{ab}\partial_\eta\;,\nonumber\\
{\cal P}_{(0)ij}^{ab}&=&\delta^{ab}\partial_i \partial_j-ig\mathcal{A}_{(0)i}^{ab}\partial_j-ig\mathcal{A}_{(0)j}^{ab}\partial_i-ig\left(\partial_i\mathcal{A}_{(0)j}\right)^{ab}-g^2\mathcal{A}_{(0)i}^{ac}\mathcal{A}_{(0)j}^{cb}\; ,\nonumber\\ 
{\cal P}_{(2)\eta\eta}^{ab}&=&-2ig\mathcal{A}_{(2)\eta}^{ab}\partial_\eta\; .
\end{eqnarray}
In deriving the above expressions we have used the fact that the background field is boost invariant, $\partial_\eta \mathcal{A}_I\equiv 0$.  It is this property of the background field that leads to the simple form of ${\cal P}_{(0)\eta i}$ and yields the factor of two in the expression\footnote{It may be instructive to work this term out explicitly.  Since the field strength tensor is anti--symmetric we have ${\cal P}_{\eta \eta}^{ab}={\cal D}_\eta^{ac} {\cal D}_\eta^{cb}$.  We can therefore write ${\cal P}_{(2)\eta\eta}^{ab}={\cal D}_{(0)\eta}^{ac} {\cal D}_{(2)\eta}^{cb}+{\cal D}_{(2)\eta}^{ac} {\cal D}_{(0)\eta}^{cb}$.  From eq.~(\ref{eq:Dexp}) we know that ${\cal D}_{(0)\eta}^{ab}=\delta^{ab}\partial_\eta$ and therefore commutes with the boost invariant background field.   The final result is then ${\cal P}_{(2)\eta\eta}^{ab}=2\mathcal{D}_{(2)\eta}^{ab}\partial_\eta$.} for ${\cal P}_{(2)\eta \eta}$. 

\subsection{Early time behavior}
In the case of the $\tau,\eta,\x_\perp$ system of coordinates and in
the Fock--Schwinger gauge, the equations of motion for the
fluctuations propagating over such a background field are written
explicitly in eqs.~(\ref{eq:linqcd}).  Solving for the full time
dependence of the fluctuations is both intractable (since the time
dependence of the background field is only known numerically) and
unnecessary (since we only need the spectrum of fluctuations at early
times).

A crucial property of the background Glasma fields is that they are
invariant under boosts in the longitudinal direction. This implies
that the fields ${\cal A}_i$ and ${\cal A}_\eta$ in the forward
light-cone, after the collision, are independent of the
space-time rapidity $\eta$. Thus the variable $\nu$, the Fourier
conjugate of $\eta$, is a conserved quantum number for fluctuations
propagating over the Glasma fields, that can be used to label the
elements of the basis.  We can therefore write the elements of the
basis as
\begin{equation}
a^\mu_{\nu \l \lambda}(\tau,\eta,\x_\perp)\equiv e^{i\nu\eta}\,
\beta_{\nu \l \lambda}^\mu(\tau,\x_\perp)\; ,
\label{eq:nu}
\end{equation}
where $\lambda=1,2$ is a polarization index and $\l$ collectively
represents the remaining quantum numbers necessary to label the
basis. The main difference with the free case (see
eq.~(\ref{eq:pol-curv})) is that we cannot assume that the $\x_\perp$
dependence of the fluctuations has the form of a plane wave and
instead represent it more generally as the function $\beta_{\nu \l
  \lambda}^\mu(\tau,\x_\perp)$. This is because the background field
has a non-trivial dependence on $\x_\perp$. A further consequence is
that, in contrast to the vacuum case, the remaining quantum numbers
encoded in $\l$ will not simply be transverse momenta. As in the free
case, we shall keep only solutions that have positive frequencies in
this basis.

We would now like to motivate one of the main results of this work; a
modified form of the linearized equation of motion
eq.~(\ref{eq:linqcd}), which captures the early time behavior of a
quantum fluctuation propagating on top of the classical background
field.  The simplest way to arrive at our result is to simply replace
all of the projectors appearing in eq.~(\ref{eq:linqcd}) with the
corresponding zeroth order Taylor coefficient from the proper time
expansion given in eq.~(\ref{eq:DPexp}).  Following this procedure
gives essentially the right result except for one subtlety.  We will
argue that we also need to include one term,
$\tau^2\mathcal{P}_{(2)\eta\eta}$, that appears at higher order in the
expansion of the projector.  The resulting small--time linearized
equations of motion are,
\begin{eqnarray}
\left(\partial_\tau \tau^{-1}\partial_\tau -\tau^{-1}\mathcal{P}_{(0)ii}\right)a_\eta+\tau^{-1}\mathcal{P}_{(0)i\eta} a_i&=&0\;, \nonumber\\
\left(\partial_\tau \tau \partial_\tau-\tau^{-1}\mathcal{P}_{(0)\eta\eta}-\tau\mathcal{P}_{(2)\eta\eta}-\tau\mathcal{P}_{(0)ii}\right)a_x+\tau^{-1}\mathcal{P}_{(0)\eta x} a_\eta+\tau \mathcal{P}_{(0)ix} a_i&=&0 \;,\nonumber\\
\left(\partial_\tau \tau \partial_\tau-\tau^{-1}\mathcal{P}_{(0)\eta\eta}-\tau\mathcal{P}_{(2)\eta\eta}-\tau\mathcal{P}_{(0)ii}\right)a_y+\tau^{-1}\mathcal{P}_{(0)\eta y} a_\eta+\tau\mathcal{P}_{(0)iy} a_i&=&0\; .
\label{eq:linqcdFrobenius} 
\end{eqnarray}
The necessity of including the term ($\mathcal{P}_{(2)\eta\eta}$) can be seen by looking at the structure of the operator acting on $a_x$ in the second equation (or equivalently acting on $a_y$ in the third equation) above; 
\begin{equation}
\left(\partial_\tau \tau \partial_\tau-\tau^{-1}\mathcal{P}_{(0)\eta\eta}-\tau\mathcal{P}_{(2)\eta\eta}-    \tau\mathcal{P}_{(0)ii}\right)a_x\;,
\end{equation} 
By examining the power counting in $\tau$ of each term in this operator, it would be inconsistent to ignore the $\tau^2$ component of $\mathcal{P}_{\eta\eta}$ which is of the same order as $\mathcal{P}_{(0)ii}$.  One could argue that both $\mathcal{P}_{(0)ii}$ and $\mathcal{P}_{(2)\eta\eta}$ are suppressed by $\tau^2$ at early times relative to $\mathcal{P}_{(0)\eta\eta}$ and therefore can both be ignored.  If we drop these terms, when we {\em turn off} the background field, we would not recover the vacuum wavefunctions and this is clearly unsatisfactory.  We therefore conclude that we need to include all projectors that are of the same order in $\tau$ as the constants appearing in the vacuum case.  By the argument presented we should also include the term $\mathcal{P}_{(2)\eta\eta}$ in order to have the correct power counting.  

A more formal way of arriving at eq.~(\ref{eq:linqcdFrobenius}) is by considering the series expansion of the small fluctuations, $a_I$. Using the method of Frobenius one finds that the leading $\tau$ behavior of the transverse components behaves as $a_i\sim \tau^{i\nu}$ while that of the longitudinal component goes as $a_\eta\sim \tau^{2+i\nu}$.  These coefficients are exactly those needed to reproduce the small $\tau$ expansion of the physical solutions found in the vacuum case.  Furthermore, if we continue to use the Frobenius method we find that only a small subset of
the Taylor Coefficients in eq.~(\ref{eq:DPexp}) are needed to determine the
lowest order coefficient in the Frobenius expansion of $a_I$.  These are precisely the Taylor coefficients that have been included in eq.~(\ref{eq:linqcdFrobenius}).   

Finally, we need to discuss why we have not included Gauss's law in eq.~(\ref{eq:linqcdFrobenius}).  The reasoning is that Gauss's law is not a dynamical equation but a constraint, and therefore
not amenable to a series expansion.  This can be seen by noting that
Gauss's law,
\begin{equation}
\mathcal{G}\equiv \tau^{-1}\mathcal{D}_\eta \mathcal{E}^\eta+\tau \mathcal{D}_i \mathcal{E}^i=0
\end{equation}
is a constant of motion ($\partial_\tau{\mathcal{G}}=0$) and
therefore if Gauss's law is obeyed at $\tau=0^+$ it will remain
satisfied for all times.

\subsection{First physical solution}

For the first physical solution, we take $a_{\nu\l1}^\eta=0$ as was
done in the vacuum case.  With this choice, the first equation in
(\ref{eq:linqcdFrobenius}) coincides with Gauss's law.  The last two
equations control the evolution of the transverse components
$a_{\nu\l1}^i$.  Since we expect the time dependence in eq.~(\ref{eq:nu}) to enter in the same way as
in the vacuum case, we postulate that the solution at early times will
be of the form
\begin{equation}
  a_{\nu\l1}^\mu(\tau,\eta,\x_\perp)
  =\frac{\sqrt{\pi}e^{\pi\nu/2}}{2Q_{\nu\l1}} \left(\begin{array}{c}  b_{\nu\l1}^{x}(\x_\perp) \\ b_{\nu\l1}^{y}(\x_\perp)\\ 0 \end{array}\right)\,e^{i\nu\eta}\,
H_{i\nu}^{(2)}\left(Q_{\nu\l1}\tau\right)\;.
\label{eq:zeroth-order-1}
\end{equation}
Next we substitute eq.~(\ref{eq:zeroth-order-1}) into the early time
linearized equation of motion~(\ref{eq:linqcdFrobenius}).  Requiring
that eq.~(\ref{eq:zeroth-order-1}) is a solution to the equations of
motion leads to,
\begin{eqnarray}
&&
-\left[{\cal D}_{(0)y}{\cal D}_{(0)y}+{\cal P}^{(\nu)}_{(2)\eta\eta}\right]b^{x}_{\nu\l 1}+{\cal                         P}_{(0)yx}b^{y}_{\nu\l1}=Q_{\nu\l 1}^2 b^{x}_{\nu\l 1}\nonumber\\
&&-\left[{\cal D}_{(0)x}{\cal D}_{(0)x}+{\cal P}^{(\nu)}_{(2)\eta\eta}\right]b^{y}_{\nu\l 1}+{\cal                         P}_{(0)xy}b^{x}_{\nu\l1}=Q_{\nu\l 1}^2 b^{y}_{\nu\l 1}\; .
\label{eq:eigen1}
\end{eqnarray}
We have explicitly included a superscript $(\nu)$ on ${\cal
  P}^\nu_{(2)\eta\eta}$ in order to remind ourselves that the
derivative with respect to $\eta$ should be replaced by $i\nu$ when
acting on the exponential in eq.~(\ref{eq:zeroth-order-1}).  We should
also stress that the equations above only depend on the background
fields at $\tau=0^+$, which are known analytically.

Solving eqs.~(\ref{eq:eigen1}) amounts to finding the eigenvalues
$Q_{\nu\l 1}^2$ and eigenfunctions (the doublets
$(b^{x}_{\nu\l1}(\x_\perp),b^{y}_{\nu\l1}(\x_\perp))$) of an Hermitean
operator. Since this
operator is Hermitean, its spectrum is made of real eigenvalues, and
its eigenfunctions can be chosen to form an orthonormal basis,
\begin{equation}
\int d^2 \x_\perp\;
 b_{\nu\l 1}^{i*}(\x_\perp)b^i_{\nu \l^\prime 1}(\x_\perp)=\delta_{\l\l^\prime}\;.
\label{eq:ortho2}
\end{equation}
Note that the choice of normalization in
eqs.~(\ref{eq:zeroth-order-1}) and (\ref{eq:ortho2}) is such that the
inner product defined in eq.~(\ref{eq:norm}) is satisfied
\begin{equation}
\big(a_{\nu\l 1}\big|a_{\nu^\prime \l^\prime 1}\big)
=
2\pi\delta(\nu-\nu^\prime) \delta_{\l\l^\prime}\;.
\end{equation}
The operator that we need to diagonalize in eqs.~(\ref{eq:eigen1}) has
a spectrum that is twice larger than the size expected for the space
of solutions with polarization $\lambda=1$. Half of this spectrum is
incompatible with Gauss' law and must be discarded\footnote{In the
  free case, this operator is ${\cal O}_{ij}\equiv
  -{\bs\partial}_\perp^2 \delta_{ij}+\partial_i\partial_j$. It has two
  types of eigenfunctions: (i) $b^i=\partial_i\chi$, with eigenvalue
  $Q^2=0$, and (ii) $b^i=\epsilon_{ij}\partial_j \chi$, with
  eigenvalue $Q^2=k_\perp^2$ (where
  $\chi(\x_\perp)\equiv\exp(i\k_\perp\cdot\x_\perp)$). Only the second
  eigenfunction is compatible with Gauss' law $\partial_i b^i=0$.  Interestingly, one of the perturbative solutions of the classical equations of motion at large transverse momenta in the forward light-cone has an identical structure~\cite{KovneMW1}. }.

\subsection{Second physical solution}
As was the case for the first physical solution, the second physical
solution will maintain the same $\tau$ dependence at $\tau=0^+$ but
with a modified dispersion relation.  We therefore write the most
general form of the second physical solution as
\begin{equation}
  a_{\nu\l2}^\mu(\tau,\eta,\x_\perp)
  =\frac{\sqrt{\pi}e^{\pi\nu/2}}{2Q_{\nu\l2}} \left(\begin{array}{c}  \,                                           b_{\nu\l2}^{x}(\x_\perp)R_{-1, i\nu}^{(2)}\left(Q_{\nu\l2}\tau\right) \\ 
b_{\nu\l2}^{y}(\x_\perp)R_{-1, i\nu}^{(2)}\left(Q_{\nu\l2}\tau\right)\\ b_{\nu\l2}^{\eta}(\x_\perp)R_{+1, i\nu}^{(2)}\left(Q_{\nu\l2}\tau\right) \end{array}\right)\,e^{i\nu\eta}\; .
\label{eq:zeroth-order-2}
\end{equation}
Following the same procedure as for the first solution, we substitute
eq.~(\ref{eq:zeroth-order-2}) into the linearized equations of
motion given by eq.~(\ref{eq:linqcdFrobenius}).  Requiring that
eq.~(\ref{eq:zeroth-order-2}) is a solution to the equation of motion
at lowest order in $\tau$ leads to the following equations for
$b^{i}_{\nu\l2}$
\begin{eqnarray}
b^{x}_{\nu\l2}&=&i\nu{\cal D}_{(0)x} b^{\eta}_{\nu\l2}\nonumber\\
b^{y}_{\nu\l2}&=&i\nu{\cal D}_{(0)y} b^{\eta}_{\nu\l2}\nonumber\\
-{\cal P}_{(0)ii} b_{\nu\l 2}^{\eta}(x_\perp) &=& Q_{\nu\l2}^2 b_{\nu\l 2}^{\eta}(x_\perp)\; .
\label{eq:eigen2}
\end{eqnarray}
The first two of these equations simply give $b^{x,y}_{\nu\l2}$ in
terms of $b_{\nu\l 2}^{\eta}$. The third equation determines $b_{\nu\l
  2}^{\eta}$ as an eigenfunction of the operator $-{\cal P}_{(0)ii}$,
with eigenvalue $Q_{\nu\l2}^2$. Because this operator is Hermitean,
these eigenvalues are real, and the eigenfunctions are mutually
orthogonal,
\begin{equation}
\int d^2 \x_\perp\; b_{\nu\l 2}^{\eta*}(\x_\perp)b^\eta_{\nu \l^\prime 2}(\x_\perp)=\delta_{\l\l^\prime}\,.
\label{eq:ortho3}
\end{equation}

We can therefore write the final form of the second physical solution
in terms of the single eigenfunction $b^{\eta}_{\nu\l2}$
\begin{equation}
  a_{\nu\l2}^\mu(\tau,\eta,\x_\perp)
  =\frac{\sqrt{\pi}e^{\pi\nu/2}}{2Q_{\nu\l2}} \left(\begin{array}{c}  \,                                           i\nu R_{-1, i\nu}^{(2)}\left(Q_{\nu\l2}\tau\right){\cal D}_{(0)x} \\ 
i\nu R_{-1, i\nu}^{(2)}\left(Q_{\nu\l2}\tau\right){\cal D}_{(0)y}\\ 
\;\;\;\;R_{+1, i\nu}^{(2)}\left(Q_{\nu\l2}\tau\right) \end{array}\right)\,b_{\nu\l 2}^{\eta}(x_\perp) e^{i\nu\eta}\,,
\label{eq:zeroth-order-22}
\end{equation}
where $b_{\nu\l 2}^{\eta}(x_\perp)$ is a solution to the eigenvalue
equation~(\ref{eq:eigen2}).  Finally, let us rewrite the solution
using the small time approximation of $R^{(2)}_{\pm1,i\nu}$, as done in
the vacuum case (see eq.~(\ref{eq:zeroth-order-2-free-asy}))
\begin{equation}
a_{\bfk2}^\mu(\bfx)
\approx
\frac{\sqrt{\pi}e^{\pi\nu/2}}{2Q_{\nu\l2}}
\left(\begin{array}{c}  {\cal D}_{(0)x}\\ {\cal D}_{(0)y} \\ -(Q_{\nu\l2}\tau)^2/(\nu+2i) \end{array}\right)\,b_{\nu\l 2}^{\eta}(x_\perp)e^{i\nu\eta}
H_{i\nu}^{(2)}(Q_{\nu\l2}\tau)\; .
\label{eq:zeroth-order-2-asy}
\end{equation}

\section{Outline of an algorithm for numerical computations}
\label{sec:algo}
The results of the previous section provide all the ingredients we
need in order to evaluate an inclusive quantity such as the
energy-momentum tensor, resumming in the calculation both the large
logs of $1/x_{1,2}$ and the secular terms that plague fixed order
calculations. The algorithm for performing such a calculation can be
broken down into several independent steps:
\begin{itemize}
\item[{\bf i.}] Solve the JIMWLK equation.
  \begin{itemize}
    \item[{\bf i.a}] Generate an ensemble of color source densities
      $\rho_a(\x_\perp)$ (or, equivalently, of Wilson lines
      $\Omega_{ab}(\x_\perp)$) that represent the distribution
      $W[\rho]$ at large $x$, close to the fragmentation region of a nucleus.
    \item[{\bf i.b}] For each of these configurations, evolve it to smaller
      $x$ by using the Langevin formulation of the JIMWLK
      equation. This amounts to performing a random walk on the space
      of mappings from ${\mathbbm R}^2$ to the group SU(3)~\cite{BlaizIW1,RummuW1}.
  \end{itemize}

\item[{\bf ii.}] Pick two elements (one for each nucleus) in the above
  ensembles for each projectile, evolved at the values of $x$ relevant
  to the observable of interest. Compute the gauge fields and their
  first time derivatives on the initial surface $\tau=0^+$ immediately
  after the collision~\cite{KrasnV1,Lappi6}.

\item[{\bf iii.}] Generate fluctuations on top of the classical Glasma  fields.

\begin{itemize}
\item[{\bf iii.a}] Solve the eigenvalue equations in 
  eqs.~(\ref{eq:eigen1}) and (\ref{eq:eigen2}). In the former case,
  only the solutions that fulfill Gauss's law should be kept. It should be
  noted that in a lattice discretization, this amounts to
  diagonalizing large but sparse matrices.

\item[{\bf iii.b}] Evaluate the Hankel functions $H^{(2)}_{i\nu}$ and
  the hypergeometric functions $R^{(2)}_{\pm1,i\nu}$ at the initial
  time of interest. (Note that because they oscillate as $\ln(\tau)\to-\infty$,
  this initial time cannot be exactly zero.) To do this, one can go
  back to their defining differential equations, and solve them
  numerically, keeping only the solution that becomes a positive
  frequency plane wave as $\tau\to+\infty$.

\item[{\bf iii.c}] Superimpose on top of the classical Glasma fields
  (obtained in step {\bf ii}), a linear combination of the small
  fluctuations that are obtained by solving the eigenvalue equations
  in eqs.~(\ref{eq:zeroth-order-1}) and (\ref{eq:zeroth-order-22}),
  multiplied by random Gaussian coefficients that have a
  variance\footnote{Note that the (arbitrary) factor $N_{_K}$ in this
    variance is cancelled by the normalization of the eigenfunctions
    $a_{_K}$ and of the integration measure $d\mu_{_K}$. Indeed, in
    eq.~(\ref{eq:random1}) one has $c_{_K}\sim \sqrt{N_{_K}}$,
    $a_{_K}\sim \sqrt{N_{_K}}$ and $d\mu_{_K}\sim 1/N_{_K}$.} given by
  eq.~(\ref{eq:random2}).
\end{itemize}

\item[{\bf iv.}] For each initial condition generated in this way,
  solve numerically the classical Yang-Mills equations forward in
  time, up to the proper time at which the observable should be
  evaluated. Repeat steps {\bf iii} and {\bf iv} in order to do a
  Monte-Carlo evaluation of the average over the fluctuations of the
  initial gauge fields.

\end{itemize}

\section{Applications and Outlook}
\label{sec:applications}
With this work, we have completed the resummation of all the leading
contributions from quantum fluctuations to inclusive quantities at
early times, in the Color Glass Condensate effective field theory
approach to high energy nucleus-nucleus collisions. These quantum
fluctuations can be factorized into $\nu=0$ and $\nu \neq 0$ modes,
where $\nu$ is the Fourier conjugate to the space-time rapidity
$\eta$. The expression in eq.~(\ref{eq:fact-formula}) described the
contribution of $\nu=0$ modes, which correspond to summing the leading
logarithmic $\alpha_s\ln (1/x_{1,2})$ contributions. Our final
expression for the energy-momentum tensor, extending the expression in
eq.~(\ref{eq:fact-formula}) to include the leading secular terms, is
\begin{eqnarray}
\langle T^{\mu\nu}\rangle_{\small{\rm LLx + LInst.}} &=& \int [D\rho_1 D\rho_2]\;
W_{x_1}[\rho_1]\, W_{x_2}[\rho_2]\nonumber \\
&\times& \int \!\! \big[{\cal D}\alpha\big]\,
F_0\big[\alpha\big]\;\; T_{_{\rm LO}}^{\mu\nu} [{\mathbbm A}[\rho_1,\rho_2] + \alpha] (x)\; ,
\label{eq:final-formula}
\end{eqnarray}
where the weight functionals $W_{x_{1,2}}[\rho_{1,2}]$ satisfy the
JIMWLK renormalization group equation in eq.~(\ref{eq:H-JIMWLK1}). The
argument ${\mathbbm A}\equiv ({\cal A},{\cal E})$ denotes collectively
the components of the classical fields and their canonicallly
conjugate momenta on the initial proper time surface.  These
quantities are functionals of $\rho_1$,$\rho_2$ and, as discussed
previously, analytical expressions for these are available at
$\tau=0^+$~\cite{KovneMW1,KrasnV3,Lappi4,Lappi6}.  The initial
spectrum of fluctuations $F_0\big[\alpha\big]$, defined in
eq.~(\ref{eq:Finit-def}), requires that one compute small fluctuations
around the classical background fields ${\mathbbm A}$ as
$\tau\rightarrow 0^+$. The formal expressions for these were derived
in section \ref{sec:glasma}. In section \ref{sec:algo}, we have
outlined a practical algorithm to compute the path integral over small
fluctuations. As numerical algorithms for solving the JIMWLK equation
are now also available~\cite{BlaizIW1} and have been successfully
implemented~\cite{RummuW1}, a full fledged numerical computation of
eq.~(\ref{eq:final-formula}) is feasible in the near future. We should
emphasize that this equation is valid for any inclusive quantity, not
just the energy-momentum tensor.

There are several applications of this formalism to understand key
features of early time dynamics in heavy ion collisions. We shall
discuss a few of these here.
\begin{itemize}

\item {\it Early thermalization} It is important to emphasize that a
  numerical simulation of eq.~(\ref{eq:final-formula}) would describe
  the real time evolution of a quantum field theory that goes far
  beyond purely classical contributions but includes as well important
  quantum effects to all orders in perturbation theory. (For similar
  considerations of the relative roles of classical versus quantum
  effects within the 2PI framework, we refer readers to
  refs.~\cite{BergeR1,BergeS1,BergeBS1} and references therein.) In
  ref.~\cite{DusliEGV1}, we developed the corresponding formalism for
  a scalar $\phi^4$ theory, and demonstrated that the system developed
  an equation of state, allowing one to write the energy-momentum
  conservation equation for the resummed energy momentum tensor
  $T^{\mu\nu}$ as the closed form set of equations corresponding to
  the ideal hydrodynamics satisfied by a perfect relativistic fluid.
  We interpreted this result as arising from a phase decoherence of
  the classical trajectories in eq.~(\ref{eq:final-formula}), with the
  different initial conditions given by the spectrum of initial
  quantum fluctuations. As discussed previously, we expect the same to
  occur in QCD as well.

  It is widely believed that the strong hydrodynamic behavior seen in
  heavy ion collisions requires early thermalization. However, as
  shown for the scalar theory, and may likely also be true for gauge
  theories, this is not a necessary condition for (nearly) perfect
  fluidity. It is interesting to ask what is the proper condition for
  thermalization in a quantum field theory. One such criterion is
  based on Berry's conjecture~\cite{Berry1,Deuts1,Jarzy1,RigolDO1},
  that states that in a quantum system whose classical counterpart is
  chaotic, high lying energy eigenfunctions look like Gaussian random
  functions.  It was later argued by Srednicki~\cite{Sredn1} that such
  eigenstates would appear to be thermal, for example, if one measures the
  single particle distribution. Since the underlying classical
  Yang-Mills theory is believed to be
  chaotic~\cite{BiroGMT1,HeinzHLMM1}, the quantum system described by
  eq.~(\ref{eq:final-formula}) is a good candidate to explore these
  ideas in a quantum field theory. In our approach, the initial state
  is not an energy eigenstate, but rather a coherent state. It is 
  formulated as a sum of plane waves with random Gaussian coefficients (see eqs.~(\ref{eq:random1}) and (\ref{eq:random2}))
 but there is no constraint that restrict the Gaussian random eigenstates to states of a given energy, in contrast to 
  the states postulated by Berry. The self-interactions of the
  fields lead to a loss of this initial coherence, and it would be
  very interesting to see if thermalization occurs on the same
  time-scales as decoherence.  

  Alternatively, one can look at the spectral function, defined in
  terms of the imaginary part of the retarded Green function, for the
  appearance of quasi-particle behavior, which allows for a kinetic
  theory description in terms of single particle distributions.  It is
  interesting to explore whether there is a region of overlap between
  a description in terms of high occupation number fields and a
  kinetic theory description in terms of
  quasi-particles~\cite{MuellS1,Jeon3}. Such a regime may equivalently
  be described by a coupled set of equations for the classical fields
  and quasi-particle modes~\cite{GelisJV1} which is reminiscent of the
  description of superfluids in condensed matter physics.  Indeed, it
  is conceivable that the best description of such an overpopulated
  system, for intermediate times, might be as a Bose-Einstein
  superfluid~\cite{BlaizGLMV1}, before the inelastic processes in the
  Glasma begin to dominate~\cite{BaierMSS2,MuellSW2} and lead
  eventually to a conventional kinetic description.

  While the mechanism for thermalization is non-perturbative, it is
  still unclear whether the mechanism is weak coupling or strong
  coupling. One might anticipate that the former is more likely at
  higher energies due to the increasing dominance of semi-hard scales,
  but where such a transition occurs is unknown. It is intriguing that
  while the the mechanism of thermalization in AdS/CFT inspired strong
  coupling approaches appears very different (more ``top down" than
  ``bottom up"), there appear to be technical similarities between
  aspects of our weak coupling approach and these
  approaches~\cite{CaronCT1,BalasBBCC1}.

\item {\it Sphaleron transitions}

  As emphasized, our master formula in eq.~(\ref{eq:final-formula}) is
  valid not only for the expectation value of the energy-momentum
  tensor but also for other inclusive quantities. One such example is
  the sphaleron transition rate $\Gamma_{\rm sphal.}$, which controls
  the mean squared change in the axial charge in thermal equilibrium
  through the relation
\begin{equation}
\langle (\Delta Q_{A,q})^2\rangle_{\rm therm.} = 4 V t \Gamma_{\rm sphal.}\,,
\end{equation}
where $V$ is the spatial volume and $q$ denotes quark flavor. In
ref.~\cite{KharzKV1}, it was noted that sphaleron transitions are not
allowed in the boost invariant classical Glasma. Because the classical
dynamics is effectively 2+1-dimensional, the second homotopy group of
SU(3) gauge theory is zero, disallowing integer valued topological
transitions. The quantum fluctuations we have been describing are no
longer boost invariant, thereby allowing sphaleron transitions to go.
In the non-equilibrium Glasma, the relevant quantity for computing the
mean square change in the axial charge is the Wightman
function~\cite{MooreT1}
\begin{equation}
G_{F{\tilde F}}^{>}(X,Y)  = \left<
 {g^2\over 32 \pi^2} F_{\mu\nu}^a {\tilde F}_b^{\mu\nu}(X)\; {g^2\over 32 \pi^2} F_{\alpha\beta}^a {\tilde F}_b^{\alpha\beta}(Y)\right>\; ,
\end{equation}
with ${\tilde F}^{\mu\nu} \equiv
\frac{1}{2}\epsilon^{\mu\nu\alpha\beta}F_{\alpha\beta}$. The sphaleron
transition rate is defined in terms of this quantity as
\begin{equation}
\Gamma_{\rm sphal} = \int d^4 X\; G_{F{\tilde F}}^{>}(X,0) \; .
\end{equation}
This quantity may be computed by a formula similar to
eq.~(\ref{eq:final-formula}). It has aroused much interest recently
because in semi-peripheral heavy ion collisions, the combined effect
of large external electromagnetic ${\vec B}$ fields and a large rate
for topological transitions can lead to an induced charge separation
phenomenon, called the {\sl Chiral Magnetic effect}~\cite{KharzMW1}, with
observable consequences. Our approach allows in principle an {\it ab
  initio} computation of this effect.

\item {\it Jet quenching}

  Besides the large flow observed in heavy ion collisions, the
  apparent strong modification of rare final states (such as jets,
  high $p_\perp$ hadrons and heavy flavors) by the medium is adduced
  as confirmation of the high degree of opacity in the medium
  consistent with a strongly correlated fluid. The standard energy
  loss mechanism is radiative energy loss, which is dominated by
  collinear splittings that are primarily influenced by late time
  dynamics in a strongly interacting quark-gluon plasma. For a
  sampling of reviews, see
  refs.~\cite{BaierSZ1,Wiede1,MajumL1,GyulaVWZ1}. There are however
  potentially large modifications of the spectra of hard final states
  from energy loss at early times, that are are not included in this
  energy loss scenario~\cite{ShuryZ1}; for a recent treatment of large
  angle contributions in a framework similar to ours, see
  ref.~\cite{MehtaST1}. However, no computation has thus far included
  next-to-leading order corrections to, for example, the single
  inclusive parton spectrum at early times taking into account both
  small x evolution and multiple scattering effects. This is done in
  our formalism when $T^{\mu\nu}$ in eq.~(\ref{eq:final-formula}) is
  replaced by $E_p {dN\over d^3 {\bf p}}$; in fact,
  eq.~(\ref{eq:final-formula}) sums up a class of all order graphs
  that correspond to coherent multiple emissions where the momenta of
  all but one of the final state gluons is integrated over. A typical
  contribution is illustrated in the figure \ref{fig:jets}. 
  \begin{figure}[htbp]
    \begin{center}
      \resizebox*{8cm}{!}{\includegraphics{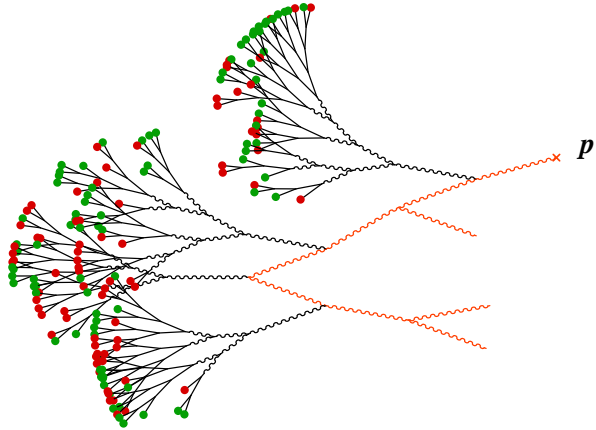}}
    \end{center}
    \caption{\label{fig:jets} Example of graph included via the
      resummation of eq.~(\ref{eq:final-formula}). The green and red
      dots represent the color charges that describe the gluon content
      of the colliding nuclei in the CGC framework.}
  \end{figure}
  Note that this diagram, corresponding to the resummed case, looks
  like a parton shower interacting with the background Glasma fields.
  However, unlike vacuum showers which can in space-time be visualized
  as being logarithmically divergent in the proper time, these ``showers"  are a consequence of the exponentially
  growing contributions from leading instabilities at each order in perturbation theory.

  We note that the single inclusive gluon spectrum also potentially
  suffers from collinear and infrared singularities. In the Glasma, it
  is possible that these could be regulated by strong multiple
  scattering and/or screening effects, but that remains to be
  proved. To avoid such complications, as in usual jet
  physics~\cite{StermW1,BashaBEL1}, one can look instead at
  correlators of the energy-momentum tensor that correspond to energy
  flow and are manifestly infrared and collinear safe. Clearly, there
  are a number of issues that need to be resolved here; the promising
  aspect of our formalism is that parton evolution, radiation and
  re-scattering can likely be treated, without ad hoc assumptions, in
  a consistent manner at early times.

\end{itemize}

\section*{Acknowledgements}
We would like to thank Miklos Gyulassy for providing the encouragement
to initiate this work. We would also like to thank T. Epelbaum,
K. Fukushima, Y. Hatta, C. Jarzynski, T. Lappi, J. Liao, L. McLerran,
A. Mueller, S. Srednyak, D. Teaney and G. Torrieri for useful
discussions. R.V. is supported by the US Department of Energy under
DOE Contract DE-AC02-98CH10886.  K.D. is supported by the US
Department of Energy under DOE Contracts DE-FG02-03ER41260 and
DE-AC02-98CH10886. F. G. would like to thank the Nuclear Theory group
at BNL for hospitality and support during the completion of this work.

\appendix

\section{Alternative basis for the free field solution}
\label{app:theta}
We will show in this appendix, that the form
of the vacuum fluctuations can be written in terms of simple analytic
functions if we trade the index $\nu$ for an index $\theta$ introduced
via the following transformation
\begin{equation}
a_{\overline\bfk1}^\mu(\bfx)
\equiv
\int\frac{d\nu}{2\pi}\;e^{-i\nu\theta}\;a_{\bfk}^\mu(\x)\; ,
\label{eq:theta-trans}
\end{equation}
where we denote $\overline\bfk\equiv(\theta,\k_\perp)$.

To see that we are free to make this change in the quantum numbers,
let us start from the real--valued field fluctuation as written in
terms of the $\nu$ coordinate \begin{equation*}
  a^\mu(x)=\sum_{\lambda=1,2}\int \frac{d^2\bfk_\perp
    d\nu}{(2\pi)^3}\;\Big[c_{\bfk\lambda}\,a_{\bfk\lambda}^\mu(x)+c_{\bfk\lambda}^*\,a_{\bfk\lambda}^{\mu*}(x)\Big]\;
  .
\end{equation*} 
Then, making the trade from $\nu$ to $\theta$ we obtain 
\begin{eqnarray*}
  a^\mu(x)&=&\sum_{\lambda=1,2}\int \frac{d^2\bfk_\perp d\nu}{(2\pi)^3}\;\Big[c_{\bfk\lambda}\,\int d\theta e^{i\nu\theta} a_{\overline{\bfk}\lambda}^\mu(x)+c_{\bfk\lambda}^*\,\int d\theta e^{i\nu\theta} a_{\overline{\bfk}\lambda}^{\mu*}(x)\Big]\nonumber\\
  &=&\sum_{\lambda=1,2}\int \frac{d^2\bfk_\perp d\theta}{(2\pi)^3}\;\Big[\int d\nu e^{i\nu\theta} c_{\bfk\lambda}\, a_{\overline{\bfk}\lambda}^\mu(x)+\int d\nu e^{i\nu\theta} c_{\bfk\lambda}^*\, a_{\overline{\bfk}\lambda}^{\mu*}(x)\Big]\nonumber\\
  &=& \sum_{\lambda=1,2}\int \frac{d^2\bfk_\perp d\theta}{(2\pi)^3}\;\Big[d_{\overline{\bfk}\lambda}\, a_{\overline{\bfk}\lambda}^\mu(x)+d_{\overline{\bfk}\lambda}^*\, a_{\overline{\bfk}\lambda}^{\mu*}(x)\Big]\; ,
\end{eqnarray*} 
where we defined 
\begin{equation*}
d_{\overline{\bfk}\lambda}\equiv\int d\nu c_{\bfk\lambda} e^{i\nu\theta}
\end{equation*} 
and in keeping with the notation employed throughout the text we have
used $\bfk\equiv(\nu,{\bf k}_\perp)$ and
$\overline{\bfk}\equiv(\theta,{\bf k}_\perp)$.  Since
$c_{\bfk\lambda}$ is a Gaussian random variable so is its Fourier
Transform, $d_{\overline{\bfk}\lambda}$, and we are free to generate
random fluctuations using either basis.

Using this new basis, the first physical solutions transforms to
\begin{equation}
a_{\overline\bfk1}^\mu(\bfx)
\equiv
\int\frac{d\nu}{2\pi}\;e^{-i\nu\theta}\;a_{\bfk1}^\mu(\x)
=
\frac{i}{2 \sqrt{\pi}k_\perp}
\left(\begin{array}{c} \k_y \\ -\k_x \\ 0 \end{array}\right)\;
e^{-ik_\perp\tau\cosh(\theta-\eta)+i\k_\perp\cdot\x_\perp}
\; ,
\end{equation}
After the transformation to the $\theta$ variable,
eqs.~(\ref{eq:soln2deint}) can be integrated over time
\begin{eqnarray}
\alpha_{2\perp}&=&
\frac{i}{2\sqrt{\pi}k_\perp} \tanh(\theta-\eta)\,e^{-ik_\perp\tau\cosh(\theta-\eta)}
\nonumber\\
\alpha_{2\eta}&=&
\frac{i}{2\sqrt{\pi}k_\perp}
\left[\frac{1+ik_\perp\tau\cosh(\theta-\eta)}{\cosh^2(\theta-\eta)}\right]
\,e^{-ik_\perp\tau\cosh(\theta-\eta)}\; .
\end{eqnarray}

Note that eqs.~(\ref{eq:soln2de}) only specify $\alpha_{2\perp}$ and
$\alpha_{2\eta}$ up to a $\tau$ independent function.  The $\tau$
independent functions that can be added to $\alpha_{2\perp}$ and
$\alpha_{2\eta}$ are not independent, because this modification must
correspond to a residual gauge transformation. Thus the allowed
modifications are
\begin{equation}
\alpha_{2\perp} \to \alpha_{2\perp}+f\quad,\qquad
\alpha_{2\eta}\to \alpha_{2\eta}+i(\partial_\theta f)\; ,
\end{equation}
where $f$ is an arbitrary $\tau$ independent function.  We can choose to
fix this residual gauge freedom by choosing the function $f$ such that
$a_{\overline\bfk2}^\mu(\bfx)$ vanishes at $\tau=0^+$.  After fixing the residual gauge freedom, we have the following
expression for the second physical solution
\begin{equation}
a_{\overline\bfk2}^\mu(\bfx)
=
\frac{i}{2\sqrt{\pi}k_\perp}
\left(\begin{array}{c} \k_x g_\perp\\ \k_y g_\perp \\ \spc\spc\spc g_\eta \end{array}\right)
\,
e^{-ik_\perp\tau\cosh(\theta-\eta)+i\k_\perp\cdot\x_\perp}\; ,
\label{eq:freesoln2}
\end{equation}
where we denote,
\begin{eqnarray}
g_\perp
&=&i\tanh(\theta-\eta)\left[1-e^{i k_\perp\tau\cosh(\theta-\eta)}\right]\,\nonumber\\
g_\eta
&=&
-\frac{1-e^{i k_\perp\tau\cosh(\theta-\eta)} 
+ i k_\perp\tau \cosh(\theta-\eta)}{\cosh^2(\theta-\eta)}\; .
\end{eqnarray}

\section{Wightman functions for free fields}
\label{sec:gamma2}
In this appendix, we shall construct the equal $\tau$ Wightman function for
free fields. 
The equal-time Wightman function
corresponding to the first physical solution
$a_{\overline{\bfk}1}^\mu$ is given by
\begin{equation}
G^{\mu\nu}_{1}
\equiv
\int\frac{d^2\k_\perp}{(2\pi)^2}d\theta\;
a_{\overline\bfk1}^{\mu*}(\tau,\eta,\x_\perp)
a_{\overline\bfk1}^\nu(\tau,\eta^\prime,\x_\perp^\prime) \; .
\end{equation}
Using the explicit expression in eq.~(\ref{eq:theta-trans}),
$G^{\mu\nu}_{1}$ can be written in the following formal way,
\begin{equation}
G^{ij}_{1}=
\frac{1}{(2\pi)^2}
\left(\begin{array}{cc} -\partial_y^2 & \partial_y\partial_x \\ 
\partial_x\partial_y & -\partial_x^2     \end{array}\right)
\mathcal{F}_{1}\left(i\partial_\eta,2\sqrt{-\tau^2\partial_\perp^2}\right)
\delta\left(\eta-\eta^\prime\right)
\ln\left(\frac{1}{\Lambda\vert x_\perp -x_\perp^\prime\vert}\right)\;,
\end{equation}
where $\Lambda$ is an infrared cutoff\footnote{The final result does
  not depend on this cutoff, thanks to the derivatives acting on the
  logarithm.} and we have defined
\begin{equation}
\mathcal{F}_{1}(\nu,d_\perp)
\equiv
\frac{1}{2}\int  \frac{dx}{\sqrt{1+x^2}}\,d\eta\;
 e^{i\nu\eta}\, e^{-ix d_\perp\sinh\left(\eta/2\right)}\; .
\label{eq:F1-def}
\end{equation}
Likewise, the Wightman function corresponding to the second physical
solution is given by
\begin{equation}
G^{\mu\nu}_{2}\equiv
\int\frac{d^2\k_\perp}{(2\pi)^2}d\theta\;
a_{\overline\bfk2}^{\mu*}(\tau,\eta,\x_\perp)
a_{\overline\bfk2}^\nu(\tau,\eta^\prime,\x_\perp^\prime) \; .
\end{equation}
From the discussion following eq.~(\ref{eq:soln2de}), we noted that
there was a residual gauge freedom remaining even after finding the
two physical solutions.  In the appendix \ref{app:theta}, we chose to
fix this gauge by requiring that $\alpha_\eta\to 0$ at $\tau=0^+$ in
order that Hamilton's equations are regular at $\tau=0$.  Clearly, for
the problem at hand, this is the correct choice.  However, computing
the Wightman function will be made much easier by choosing a different
gauge,
\begin{equation}
g_\perp=i\tanh(\theta-\eta)\nonumber\quad,\qquad
g_\eta=-\frac{1+ik_\perp\tau\cosh(\theta-\eta)}{\cosh^2(\theta-\eta)}\; .
\label{eq:gperp-1}
\end{equation}
We should stress that there is a residual gauge freedom remaining in
any correlator of gauge fields.  With this new gauge choice we find
\begin{equation}
G^{ij}_{2}
=
-
\frac{1}{(2\pi)^2}
\left(\begin{array}{cc} \partial_x^2 & \partial_y\partial_x \\ \partial_x\partial_y & \partial_y^2     \end{array}\right)
\mathcal{F}_{2}\left(i\partial_\eta,2\sqrt{-\tau^2\partial_\perp^2}\right)
\delta\left(\eta-\eta^\prime\right)
\ln\left(\frac{1}{\Lambda\vert x_\perp -x_\perp^\prime\vert}\right)\; ,
\end{equation}
with
\begin{equation}
\mathcal{F}_{2}(\nu,d_\perp)
\equiv
\frac{1}{2}\int \frac{dx}{\sqrt{1+x^2}}\,d\eta\;
\left[1-\frac{2\cosh \eta}{1+2x^2+\cosh \eta}\right]
e^{i\nu\eta}\, e^{-ix d_\perp\sinh\left(\eta/2\right)}\; .
\label{eq:F2-def}
\end{equation}

We can also calculate the correlation function of the canonical momenta
defined by $e^i=\tau \partial_\tau a_i$.  The free Wightman function
for the transverse components of the electric field is defined as
\begin{equation}
H^{ij}
=
\sum_{\lambda=1,2} \tau^2 
\int\frac{d^2\k_\perp}{(2\pi)^2}d\theta\;
\Big[\partial_\tau a_{\overline\bfk\lambda}^{i*}(\tau,\eta,\x_\perp)\Big]\,
\Big[\partial_\tau a_{\overline\bfk\lambda}^j(\tau,\eta^\prime,\x_\perp^\prime)\Big]\; . 
\end{equation}
Note that the $\tau$ derivatives remove the residual gauge freedom
that still remained in the expression for $G^{ij}$ as written above.
This is expected since we don't expect any residual gauge freedom to
remain when computing correlators of physical quantities such as th
electric field.  The Wightman function for the first physical solution
is
\begin{equation}
H^{ij}_{1}=
\frac{1}{(2\pi)^2}
\left(\begin{array}{cc} -\partial_y^2 & \partial_y\partial_x \\ 
\partial_x\partial_y & -\partial_x^2     \end{array}\right)
\mathcal{F}_{3}\left(i\partial_\eta,2\sqrt{-\tau^2\partial_\perp^2}\right)
\delta\left(\eta-\eta^\prime\right)
\ln\left(\frac{1}{\Lambda\vert x_\perp -x_\perp^\prime\vert}\right)\;,
\end{equation}
and similarly for the second physical solution
\begin{equation}
H^{ij}_{2}
=
-
\frac{1}{(2\pi)^2}
\left(\begin{array}{cc} \partial_x^2 & \partial_y\partial_x \\ \partial_x\partial_y & \partial_y^2     \end{array}\right)
\mathcal{F}_{4}\left(i\partial_\eta,2\sqrt{-\tau^2\partial_\perp^2}\right)
\delta\left(\eta-\eta^\prime\right)
\ln\left(\frac{1}{\Lambda\vert x_\perp -x_\perp^\prime\vert}\right)\; ,
\end{equation}
where the functions $\mathcal{F}_{3,4}$ are defined as
\begin{equation}
\mathcal{F}_{3}(\nu,d_\perp)
\equiv
\frac{1}{4}\int \frac{dx}{\sqrt{1+x^2}}\,d\eta\;
\left[1+2x^2+\cosh \eta\right]
e^{i\nu\eta}\, e^{-ix d_\perp\sinh\left(\eta/2\right)}\; ,
\label{eq:F3-def}
\end{equation}
\begin{equation}
\mathcal{F}_{4}(\nu,d_\perp)
\equiv
\frac{1}{4}\int \frac{dx}{\sqrt{1+x^2}}\,d\eta\;
\left[1+2x^2-\cosh \eta\right]
e^{i\nu\eta}\, e^{-ix d_\perp\sinh\left(\eta/2\right)}\; .
\label{eq:F4-def}
\end{equation}

\end{document}